\documentclass[lettersize,journal]{IEEEtran}
\usepackage{amsmath,amsfonts}
\usepackage{algorithmic}
\usepackage{algorithm}
\usepackage{array}
\usepackage[caption=false,font=normalsize,labelfont=sf,textfont=sf]{subfig}
\usepackage{textcomp}
\usepackage{stfloats}
\usepackage{url}
\usepackage{verbatim}
\usepackage{graphicx}
\usepackage{cite}
\usepackage{xcolor}
\usepackage{multirow}
\usepackage{lipsum}
\usepackage{soul}
\usepackage{pifont}
\let\subfigure\subfloat
\newcommand{\xmark}{\ding{55}} 
\begin{document}
\setstcolor{red}

\title{Assessing the Challenges of Collective Perception via V2I Communications in High-Speed Scenarios with Open Road Testing}

\author{Jon Ander Iñiguez de Gordoa, Iker Alkorta, Itziar Urbieta, Gorka Velez, and Andoni Mujika
\thanks{Jon Ander Iñiguez de Gordoa, Iker Alkorta, Itziar Urbieta and Gorka Velez are with Fundación Vicomtech, Basque Research and Technology Alliance (BRTA), Mikeletegi 57, 20009 Donostia-San Sebasti\'an, Spain (e-mail: \{jainiguez, ialcorta, iurbieta, gvelez\}@vicomtech.org).}
\thanks{Jon Ander Iñiguez de Gordoa and Andoni Mujika are with the Department of Computer Sciences and Artificial Intelligence, University of the Basque Country (UPV/EHU), Donostia-San Sebasti\'an
20018, Spain (e-mail: \{jiniguezdegord001.ikasle, andoni.mujika\}@ehu.eus).}
}


\markboth{%
    \NoCaseChange{%
        \begin{tabular}{@{}l@{}}
            This article has been accepted for publication in IEEE Transactions on Vehicular Technology. This is the author's version which has not been fully edited and \\ 
            content may change prior to final publication. Citation information: DOI 10.1109/TVT.2026.3672175
        \end{tabular}%
    }%
}{}


\IEEEpubid{DOI 10.1109/TVT.2026.3672175 ~\copyright~2026 IEEE}

\maketitle

\IEEEoverridecommandlockouts
\IEEEpubid{\begin{minipage}{\textwidth}\ \\\\\\\\\\[12pt]
\copyright 2026 IEEE. All rights reserved, including rights for text and data mining and training of artificial intelligence and similar technologies. Personal use is permitted, but republication/redistribution requires IEEE permission. See https://www.ieee.org/publications/rights/index.html for more information.
 \end{minipage}}

\begin{abstract}
This paper presents a comprehensive end-to-end evaluation of an infrastructure-assisted collective perception (ICP) system deployed on a highway using ITS-G5 technology. Open-road tests were conducted in the Bizkaia Connected Corridor (BCC), an operational corridor which covers a winding highway, enabling a realistic assessment of system performance in diverse traffic scenarios. The evaluation included three main aspects: (1) end-to-end Vehicle-to-Everything (V2X) communication latency, with a breakdown of delays introduced by each system component; (2) the effective range of ITS-G5 communications between vehicles and infrastructure; and (3) the perception system, using an independent sensor setup for ground truth annotation to account for errors beyond the detection model, such as synchronization, localization, and calibration inaccuracies. The results reveal that object detection and asynchronous transmission of collective perception messages (CPMs) are major latency bottlenecks, with results showing that synchronizing CPM transmission with local perception can reduce delays by up to 33\%. Additionally, onboard perception struggles with detecting objects beyond 50 meters, highlighting the importance of collective perception in highway environments, where communication ranges significantly exceed detection limits. The findings provide valuable insights to optimize ICP deployments, supporting safer and more efficient cooperative mobility systems.
\end{abstract}


\section{INTRODUCTION}


Vehicle-to-Everything (V2X) communication enables real-time data exchange between vehicles, infrastructure, and other road users, enhancing traffic safety and efficiency. Among its applications, Collective Perception (CP) allows vehicles and infrastructure to share sensor data, expanding situational awareness beyond line-of-sight obstacles and providing a more comprehensive view of the surrounding environment.

Connected vehicles can leverage Infrastructure-to-Vehicle (I2V) communications to overcome the limitations of coexisting with road users without V2X capabilities, such as vulnerable road users (VRUs) and legacy vehicles, which cannot broadcast their presence. Infrastructure-assisted Collective Perception (ICP) provides several key advantages: Roadside Units (RSUs) can be strategically positioned to ensure high detection availability, with their placement optimized for V2X signal propagation within the area of interest. Additionally, since RSUs are static, their absolute positions are precisely known, minimizing transformation errors. Furthermore, it allows vehicles to operate with a more streamlined or reduced sensor setup, relying on external data for enhanced perception in complex scenarios. By integrating CP, connected vehicles can enable advanced safety applications, including VRU protection, Overtaking Warning, and Advanced Intersection Collision Warning \cite{car2019guidance}.

Conversely, infrastructure can also benefit from perception data provided by vehicles (Vehicle-to-Infrastructure, V2I). Vehicles offer a dynamic perspective, complementing the static viewpoint of roadside infrastructure. This exchange enables the development of enhanced infrastructure-based safety applications, cooperative lane-changing and merging assistance (e.g., recommending optimal merging speeds), improved traffic management through real-time smart rerouting, automated incident detection (e.g., debris or stopped vehicle detection, wrong-way driver alerts), and faster emergency response coordination, ultimately improving overall road safety and efficiency.

Due to the mentioned opportunities arising from vehicle-infrastructure interactions, this paper presents a comprehensive evaluation of an ICP system in a high-speed environment, leveraging ITS-G5 communications and state-of-the-art technologies, with a special focus on V2I communications. This evaluation covers every component of the system architecture, from the V2X communication latency and range to the accuracy of the sensing vehicle's perception system.

This study analyses V2X communication performance by assessing end-to-end latencies and transmission ranges in the deployed architecture. These factors are particularly critical for CP in high-speed environments, where even small temporal delays translate into significant distances traveled by vehicles and the detected objects. The evaluation provides a detailed breakdown of the latency components and an analysis of the effective communication range within the test corridor, validating the reliability of ICP under real-world conditions and demonstrating alignment with state-of-the-art benchmarks and specifications.

\IEEEpubidadjcol
Moreover, existing perception performance benchmarks often focus on detection or detection-tracking modules only, using ground truth data annotated directly on the same sensor inputs (e.g., images or point clouds) used for the detections. This approach overlooks potential errors or misalignments introduced by other components, such as the sensor setup and calibration. This paper addresses this gap by proposing a comprehensive evaluation methodology that explicitly incorporates the effects of sensor setup and calibration, detector accuracy, synchronization, and processing delays. To ensure accurate ground truth annotation, a highly calibrated multi-sensor setup, independent of the vehicle's configuration, is employed during the open road tests. This approach provides not only empirical results but also a systematic foundation to evaluate the real-world feasibility of CP services under practical conditions.

The tests were conducted in the Bizkaia Connected Corridor\footnote{https://bizkaiaconnectedcorridor.biz/en} (BCC), an operational connected corridor, ensuring the results are grounded in real and practical deployments. This corridor is deployed on a winding highway through mountainous terrain in the Basque Country, Spain. The tests were carried out in open traffic and a high-speed environment, capturing scenarios such as toll booths, overtaking maneuvers, and varying traffic densities, providing a realistic assessment of the system’s performance under diverse real-world conditions.

While much of the existing literature relies on simulations or controlled test environments, some studies have examined vehicular communications and CP in real-world settings. However, previous research on corridors has primarily focused on cellular communications, such as long-term evolution (LTE) or 5G, rather than short-range communications, with an emphasis on the key challenge of service continuity \cite{via2022, zfernandez2023, slamnik2024, Gallego2024}.

Prior studies utilizing short-range communication technologies like ITS-G5 or dedicated short-range communications (DSRC) in open-road conditions have addressed different aspects, including physical layer (PHY) performance \cite{Moradi2023}, road design criteria \cite{Guerrieri2021}, and driving behavior \cite{Kang2023}. Almeida et al. \cite{almeida2023real} studied the generation and processing times of ETSI Collective Perception Messages (CPM) for different message sizes but did not study the end-to-end latencies or the impact on perception performance. Pilz et al. \cite{pilz2023collective} investigated the delay chain in a CP system involving two vehicles but they did not consider the roadside infrastructure, and like \cite{almeida2023real}, they also did not assess the perception performance. 

The present work is therefore novel in its comprehensive evaluation of an ITS-G5-based ICP system in an open-road corridor, providing new insights into real-world deployment and performance. More specifically, the key contributions of this work are:
\begin{itemize}
    \item We propose a novel CP evaluation methodology that integrates end-to-end communication performance and a full-stack perception analysis. Unlike conventional approaches that isolate perception or PHY-level communication, our methodology considers all components of the CP pipeline, including sensor calibration, detection, tracking, synchronization and latency decomposition. This integral evaluation not only provides a comprehensive assessment of the system's performance in the analysed scenarios, but also establishes a theoretical framework for end-to-end CP evaluation, enabling reproducibility and comparability across different deployments.
    
    \item Key insights for deploying CP services (CPS) on highways were identified, based on the results obtained from the evaluation of the perception, latencies and communication ranges. This comprehensive assessment allowed us to pinpoint critical bottlenecks and areas with the greatest potential for improvement, providing valuable guidance to optimize future implementations. These insights not only highlight current limitations but also offer concrete recommendations to enhance system performance and reliability.

\end{itemize}

The rest of the paper is organized as follows: Section \ref{sec:sota} reviews related work relevant to this study. Section \ref{sec:architecture} describes the architecture of the ICP system analyzed in this paper. Section \ref{sec:evaluation} outlines the employed evaluation methodology. Section \ref{sec:results} presents the obtained results. Finally, Section \ref{sec:conclusions} discusses the key conclusions and insights derived from this study.
\section{RELATED WORK}
\label{sec:sota}

\subsection{WIRELESS COMMUNICATIONS FOR COLLECTIVE PERCEPTION}
CP enables Connected and Automated Vehicles (CAVs) to exchange information about their surroundings, enhancing situational awareness. The fusion of this shared information is categorized into three levels \cite{Zhengwei2024}: early, intermediate, and late fusion, depending on the type of data exchanged.

In early fusion, raw sensor data from multiple agents is transmitted. This approach requires aligning sensor outputs to a common reference frame, which becomes complex when multiple moving agents are involved, and also demands significant bandwidth~\cite{yuan2022keypoints}. Intermediate fusion reduces bandwidth requirements by sharing processed features from object detectors, such as voxel~\cite{chen2019f} or spatial-temporal features \cite{zhang2024efficient}. However, this approach is only viable when using identical or compatible detection architectures. Finally, in late fusion, only high-level object data is shared, including detected objects' position, dimensions, and dynamic attributes. This method significantly lowers bandwidth demands and facilitates interoperability between different systems. For late fusion, the Collective Perception Message (CPM) standardized by ETSI can be employed \cite{etsi2019intelligent}.

The ETSI CPM specification is agnostic to the underlying communication technology and has been implemented over both IEEE 802.11p-based and cellular-based V2X (C-V2X) systems. ITS-G5, the European variant of IEEE 802.11p operating in the 5.9 GHz ITS band, has been widely adopted in real-world deployments due to its low latency and direct communication capabilities. In the case of C-V2X, messages can be transmitted via the PC5 interface for direct sidelink communication or through the Uu interface, which exploits existing cellular infrastructure \cite{hajisami2022tutorial}. When using the Uu interface, messages are usually transmitted using a publish-subscribe protocol such as Message Queuing Telemetry Transport (MQTT) or Advanced Message Queuing Protocol (AMQP) \cite{fmogollon2024}. However, using a broker to communicate between vehicles, or between a vehicle and infrastructure, adds a significant delay \cite{jarin2024}, which limits its suitability for real-time, safety-critical applications, although they remain useful for certain use cases that require long-distance communication or cloud processing. 

Infrastructure-assisted Collective Perception (ICP) has gained increasing attention as a means to enhance vehicular perception beyond the limitations of on-board sensors. This approach consists of deploying RSUs often equipped with their own sensors and computational resources, incorporating edge computing capabilities \cite{malik2023collaborative}. Recent research efforts have explored various aspects of ICP, including its advantages over vehicle-to-vehicle (V2V) collective perception and message prioritization strategies.

Several studies have demonstrated the benefits of ICP over vehicle-based approaches. Ku et al. \cite{ku2023uncertainty} demonstrated the feasibility of offloading tasks like perception fusion from vehicles to edge computing units within the infrastructure. Schiegg et al. \cite{schiegg2023accounting} highlighted the role of ICP in improving environmental perception by introducing packet duplication and prioritization mechanisms, achieving a significant increase in detection performance, particularly in low-connectivity scenarios. Similarly, Chtourou et al. \cite{chtourou2021collective} analyzed the effectiveness of RSUs in extending vehicles’ perception range, showing that ICP outperforms vehicle-based CP by up to a factor of eight in terms of perceived objects. Huang et al. \cite{huang2019performance} further evaluated the effectiveness of V2V-based CP in highway scenarios, demonstrating that even with a 10\% connected vehicle penetration rate, perception coverage increases significantly, though communication overhead challenges must be addressed.

Another crucial challenge in ICP research is optimizing communication efficiency and message dissemination. Wolff et al. \cite{wolff2023infrastructure} proposed an approach based on the Value of Information (VoI) to reduce network load while maintaining high awareness of VRUs. Similarly, Merwaday et al. \cite{merwaday2021infrastructure} introduced a layered costmap-based perception-sharing protocol to mitigate the communication overhead associated with transmitting large sets of detected objects, demonstrating significant reductions in data transmission requirements. Jia et al. \cite{jia2023mass} developed a mobility-aware sensor scheduling algorithm that optimizes the selection of sensor data from cooperative vehicles, which enhances the overall perception quality and reduces communication overhead.
Almeida et al. \cite{almeida2023real} conducted real-world tests of CP in a connected city infrastructure, evaluating the impact of message size on dissemination efficiency and highlighting the need for adaptive message filtering mechanisms. Furthermore, various studies have focused on defining CPM generation rules in order to minimize redundant data and avoid future channel congestion\cite{thandavarayan2020generation,shule2022tracking,chtourou2021context,thandavarayan2019analysis,masuda2022feature}.





    \subsection{EVALUATION OF COLLECTIVE PERCEPTION SYSTEMS}
    \label{sec:sota:eval}
Evaluating the multifaceted performance of CP systems has also been a key research focus that requires moving beyond standard perception metrics commonly used for single-agent systems. While metrics like Average Precision (AP), Intersection over Union (IoU), precision, and recall remain relevant for assessing basic detection accuracy, a broader suite of metrics is needed to capture the specific benefits and challenges of cooperation. Communication performance is fundamental to CPS, necessitating metrics such as end-to-end latency (E2E) or packet delivery rate (PDR) \cite{Asabe2024}. These communication metrics quantify the efficiency and reliability of the underlying V2X data exchange.

Pilz et al. \cite{pilz2023collective} proposed a decomposition of the CP delay chain, delineating five primary components: sensing delay, local fusion delay, communication delay, asynchronous delay, and global fusion delay. They also conducted real-world experiments with automated driving demonstrators to measure delay components in a practical CP system, providing insights into the impact of latency on the applicability of CP use cases. Similarly, Volk et al. \cite{volk2021towards} emphasized the importance of incorporating sensor-based processing delays into simulations, enabling a more accurate assessment of the end-to-end latency of the CP pipeline. 

Recent CP datasets, such as DAIR-V2X \cite{yu2022dair}, V2V4Real \cite{xu2023v2v4real}, or Tumtraf \cite{zimmer2024tumtraf}, also include real-world data recorded from different vehicles and RSUs. However, most of these datasets cover a very limited range of scenarios, as they only include data from specific locations (e.g., intersections) where RSUs were deployed, thus lacking the diversity needed to fully validate a CP system across all potential operation domains. Moreover, in these datasets, ground truth is annotated at least partially in the same sensor setup used for detection. This methodology overlooks potential errors introduced by the perception sensors themselves, leading to an overly optimistic assessment of the system performance.

Previous studies have also explored the optimization of sensor configurations through simulations or synthetic data, aiming to increase the accuracy of the perception data that would then be transmitted in CP systems. 
For instance, Ma et al. \cite{ma2021perception} introduced the perception entropy metric to measure and minimize perception uncertainty around a vehicle, and validated their solution using simulations. 
Additionally, Hu et al. \cite{hu2022investigating} analyzed the impact of multi-LIDAR placement on object detection using synthetic data, demonstrating that sensor placement might influence the performance of object detection by up to 10\%. Still, to the best of our knowledge, our work is the first to address sensor setup-introduced errors in real-world deployments, a critical factor in order to understand the true limitations of CP systems in practical applications.

\begin{table}
    \centering
        \caption{Comparison of this work (end-to-end and module-level latency, communication range, and comprehensive perception assessment) against previous studies.}
    \begin{tabular}{|l|c|c|c|c|}
    \hline
         \textbf{Study} & \textbf{Latency} & \textbf{Range} & \textbf{Perception} \\
    \hline
       Pilz et al. \cite{pilz2023collective}  &  \checkmark & \xmark & \xmark  \\
       Asabe et al. \cite{Asabe2024}  & $\sim$ & \checkmark &  \xmark  \\
       Schiegg et al. \cite{schiegg2023accounting}& \xmark & \checkmark & $\sim$ \\
       Hu et al. \cite{hu2022investigating} & \xmark & \xmark & $\sim$ \\
    \hline
         This study& \checkmark & \checkmark & \checkmark  \\
    \hline
    \end{tabular}
    \label{tab:sota_comparison}
\end{table}

Table \ref{tab:sota_comparison} summarizes how the evaluation in this work compares to previous relevant studies, highlighting the inclusion of multiple dimensions for a comprehensive assessment of a real-world CPS deployment.

Despite these advancements, there remains a gap in evaluating ICP under real-traffic conditions on highways. Most studies rely on simulations or urban test environments, whereas real-world highway scenarios introduce unique challenges related to high-speed dynamics, varying connectivity levels, and latency constraints. In order to provide sufficient time for decision-making and control, CP systems require lower latencies and longer detection ranges in high-speed scenarios than in slower urban traffic. In this work, we present an empirical evaluation of an ICP system deployed on a highway in real traffic, analyzing effective communication range, end-to-end latency across the entire pipeline and perception performance.

\section{SYSTEM ARCHITECTURE}
\label{sec:architecture}

\begin{figure*}
    \centering
\includegraphics[width=0.8\linewidth]{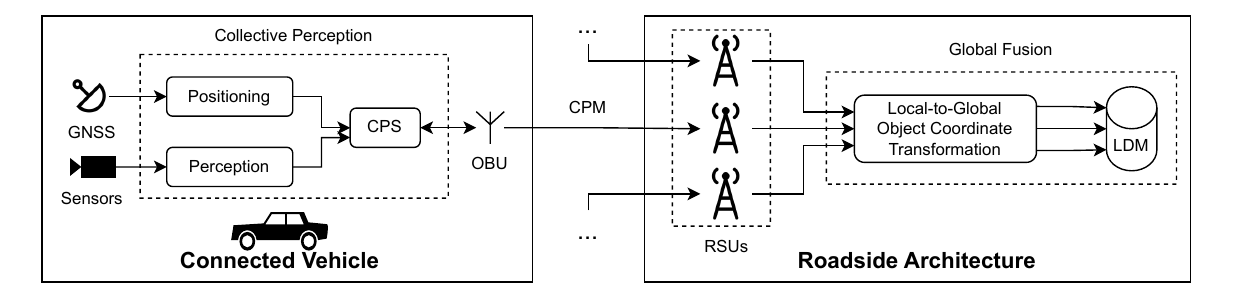}
    \caption{ICP architecture used in this study.}
    \label{fig:architecture}
\end{figure*}

Figure \ref{fig:architecture} presents a simplified diagram of the ICP architecture used in this study. Connected vehicles use their positioning and onboard perception systems to generate and transmit CPMs. These messages are received by the RSUs in the roadside infrastructure, where real-time data fusion can be performed to enhance local situational awareness. 

The roadside infrastructure is connected to a Traffic Management Center (TMC), which supports higher-level functions such as data storage, broadcasting centralized warnings, and traffic coordination. However, safety-critical applications demand ultra-low latency, particularly in high-speed scenarios where every millisecond impacts reaction time. Involving the TMC in the real-time loop would introduce additional network hops and increase end-to-end latency. Instead, edge processing at the RSU level minimizes end-to-end latency, ensuring the timely delivery of safety messages. For this reason, in this study, end-to-end communication refers specifically to the direct exchange between vehicles and roadside infrastructure. While the TMC plays a role in higher-level traffic coordination, its functions fall outside the scope of the real-time evaluation of this study. 



\subsection{CONNECTED VEHICLE}
A Toyota Prius equipped with CP capabilities was used as the connected vehicle. The car featured a Velodyne HDL-32E LiDAR (360$^{\circ}$ horizontal FoV and 41$^{\circ}$ vertical FoV) and an OxTS XNAV 550 GNSS-aided inertial navigation system. CenterPoint \cite{Yin_2021_CVPR}, trained on the NuScenes dataset \cite{Caesar_2020_CVPR}, served as the LiDAR-based 3D object detection framework, and the algorithm proposed by Montero et al. \cite{montero2023multi} was used as the tracking model. The perception model ran onboard a Nuvo-10208GC computer with an Intel Core i7-13700E processor, 64 GB of RAM, and two Nvidia 4070 Ti GPUs. CPMs were transmitted to the RSUs via a Cohda Wireless MK5 Onboard Unit (OBU), which supports CPM encoding/decoding and ITS-G5 connectivity. The perception computer and ITS-G5 OBU communicated over Ethernet via UDP sockets, exchanging information in JSON format.

During peak CPM generation, the average CPU usage was around 9.1\% in the 24-core Intel i7-13700E, while GPU utilization was approximately 20\% on one Nvidia 4070 Ti and 10\% on the second. These values are consistent with the hardware capabilities and indicate that the system was overdimensioned relative to the task, ensuring real-time operation without resource saturation and providing flexibility during development. In production CAVs, however, power and thermal constraints would likely require lower-power or embedded alternatives, which have also been shown to provide real-time performance for similar applications \cite{herrero2023pedestrian}. Nevertheless, the perception latency reported in this paper should be considered a lower bound, as it may increase in more resource-constrained, production-grade platforms.


\subsection{ROADSIDE INFRASTRUCTURE}
\label{sec:architecture:bcc}
As stated before, the roadside infrastructure used in the study is part of the BCC corridor, which is a public-private initiative to test, validate, and demonstrate Cooperative, Connected, and Autonomous Mobility technologies, as well as Intelligent and Digital Infrastructures, in a real-world scenario. The BCC is located in Bizkaia, a region in northern Spain. The complete Living Lab consists of 1,200 km of roads, but in this study, we have focused on 69 km between the towns of Ermua and Muskiz, along the A8 and AP8 highways. This segment of the corridor contains 20 Kapsch RIS-9260 RSUs (each with an RIS-9260-1A0W antenna, generally mounted between 5 and 12 meters above the road), which provide V2X communication based on the ETSI ITS-G5 standard. Due to the mountainous terrain, despite being highways, the roads have many curves and changes in elevation. These roads vary between two and four lanes depending on the section, and the maximum speed limit varies between 80 and 120 km/h, making this drive ideal for testing heterogeneous highway conditions.



At the time of the experiments, the RSUs were not equipped with infrastructure-based perception sensors and therefore did not generate CPMs based on local sensing. The RSUs receive V2X messages, such as CPM, from the connected vehicles operating within the coverage area of the roadside infrastructure. An ICP service must be capable of aggregating and fusing data from multiple sources or vehicles, centralizing this fused data to serve as input for more advanced applications. Given the latency-sensitive nature of these applications, data fusion is performed at a single edge server within the roadside infrastructure, centralizing information from all RSUs and ensuring minimal transmission delays. This can be achieved by using a Local Dynamic Map (LDM) that has direct access to the messages received at the RSU, enabling real-time data integration and dissemination.

An LDM is a structured data storage system used to maintain and manage real-time information about the surrounding environment. It integrates static, dynamic, and temporary data from multiple sources to support cooperative and autonomous driving \cite{etsi2011102}. The LDM can be deployed in the vehicle or in the infrastructure. In a similar way to \cite{mgarcia2023}, in our implementation, we centralise the information gathered from the connected vehicles by all RSUs in a single LDM instance. 

The deployed LDM implementation consists of a backend system that collects and processes V2X messages from RSUs, centralizing data for real-time analysis. Each RSU receives CPMs from connected vehicles, which are then transmitted to a common edge server within the roadside infrastructure. The server parses these messages using custom-built parsers and stores the structured data in an InfluxDB time-series database, enabling efficient storage and retrieval of high-frequency data streams.  InfluxDB includes several features to enable scaling and ensure reliable operation in high-traffic environments. 

As explained in Section \ref{sec:eval:processing}, due to technical limitations in the BCC at the time of the road tests, each RSU performed a packet dump of the received messages, and the LDM processing using each RSU log was carried out later in a separate infrastructure.

\section{EVALUATION METHODOLOGY}
\label{sec:evaluation}
In this study, two main aspects of the deployed CP architecture are evaluated: the communications and the perception system. 

First, to evaluate the CPM communications, two key aspects are measured: (1) the latencies of the CPMs, including both end-to-end latencies and the delays introduced by each module in the architecture; and (2) the communication range, which determines how far the vehicles and the RSUs of the BCC can successfully exchange data.

Secondly, the accuracy of the CPMs transmitted by the connected vehicle is assessed by evaluating its perception system. This evaluation employs a highly calibrated sensor setup, distinct from the vehicle's own setup, to account for the errors introduced not only by the 3D object detection and tracking modules but also by sensor calibration inaccuracies, GNSS positioning errors and clock synchronization discrepancies.

By evaluating latency, communication range and perception accuracy together, we obtain a comprehensive understanding of the system’s performance, capabilities and limitations. The conclusions of this study are drawn from the combined insights of these three dimensions, rather than from a single aspect, and therefore provide a more accurate reflection of the overall behavior of the deployed CP architecture.

Security is a critical aspect of any communication system evaluation; however, it falls outside the scope of this study, which focuses on the application layer of collective perception. A key reason for this omission is that, at the time of the open-road tests, the BCC infrastructure did not support Public Key Infrastructures (PKIs), which are required for securing V2X communications. Future work should incorporate security evaluations, particularly regarding authentication, message integrity, and resilience against potential attacks on the CP system.

\subsection{GROUND TRUTH GENERATION}

As previously discussed, annotating ground truth on the same sensor data used for inference fails to account for errors introduced by the sensors themselves (such as calibration and synchronization inaccuracies, limited fields of view or occlusions). Therefore, generating ground truth data with a distinct sensor setup is preferable.

This alternative sensor setup should be multi-sensor, designed to maximize the field of view coverage, with each sensor precisely calibrated and synchronized. The annotation process does not need to occur in real time, as the data will be recorded and analyzed retrospectively. Therefore, manual or semi-automatic labeling methods can be employed. 

In this study, a vehicle equipped with AVL's Dynamic Ground Truth (DGT) system\footnote{https://www.avl.com/en/testing-solutions/automated-and-connected-mobility-testing/avl-dynamic-ground-truth-system} was used as the ground truth generator. The DGT system is a highly precise, all-in-one sensor setup that captures a 360º field of view of its environment using three time-synchronized LIDARs, four high-resolution cameras and high-precision GNSS sensors. All data streams from the DGT system are recorded in a time-synchronized manner to ensure precise alignment between different sensor inputs (sensors are triggered simultaneously to ensure precise temporal alignment across the data streams). During data collection, recordings are stored on hard drives and later transferred to a data processing center, where the offline perception is executed. The annotated data is then manually validated using the original sensor recordings to ensure accuracy.

Since the output of the DGT system is used as the ground truth of this evaluation, one might assume that this setup can also not be entirely free from calibration errors or other types of inaccuracies. However, as it undergoes a far more rigorous and recent calibration process than in the connected vehicle, its accuracy is significantly higher. For instance, sensor setups in vehicles cannot be recalibrated or adjusted daily and may degrade over time, leading to increasing inaccuracies. In contrast, ground truth generators, since they are not in continuous operation, can be calibrated with extreme precision just before testing, ensuring highly accurate reference values. The DGT system is an all-in-one sensor setup where the sensors are already precalibrated relative to each other. Right before the open road tests in this study, the DGT system was precisely calibrated with respect to the vehicle itself, following the ISO 8855 standard~\cite{iso2013road}. This calibration accounted for the specific distribution of weight inside the vehicle (including passengers and any additional load) during the tests. Specifically, the DGT system reports a root mean squared error in sensor calibration of just 1 cm and 0.1 degrees across the three axes (x, y, z) and orientations (yaw, pitch, roll), respectively. 

Another challenge when using two different sensor setups is the possibility of fully occluded objects in either of the setups, which may lead to these objects being incorrectly classified as false positives or false negatives in the evaluation process. However, the DGT system features a more extensive sensor setup with a greater detection range, and its annotations are generated through a semi-automatic process with manual verification, reducing the likelihood of the connected vehicle detecting objects that the DGT system does not perceive. On the other hand, objects detected by the DGT system but occluded from the connected vehicle’s perspective reflect real-world scenarios that any connected vehicle would encounter in traffic. Therefore, given these considerations, the DGT system can be reasonably considered a reliable ground truth reference for this evaluation.

Stationary systems for ground truth generation are not suitable for this study, as their limited coverage and differing viewpoints compared to the vehicle can create occlusions visible in one setup but not in the other, generating false positives and negatives that bias perception metrics. Using a dynamic ground truth recorder near the evaluated vehicle provides broader coverage and a perspective aligned with the vehicle, enabling a fairer assessment of its perception performance.

\subsection{EXPERIMENT SETUP}

\begin{figure}
    \centering
    \subfigure{\includegraphics[width=\linewidth]{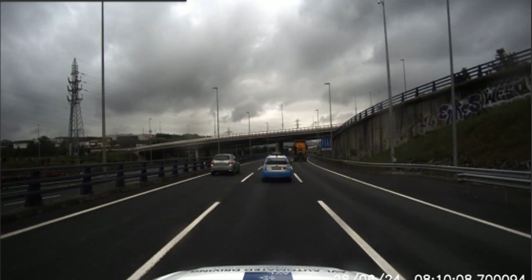}}
\hfil
\subfigure
{\includegraphics[width=\linewidth]{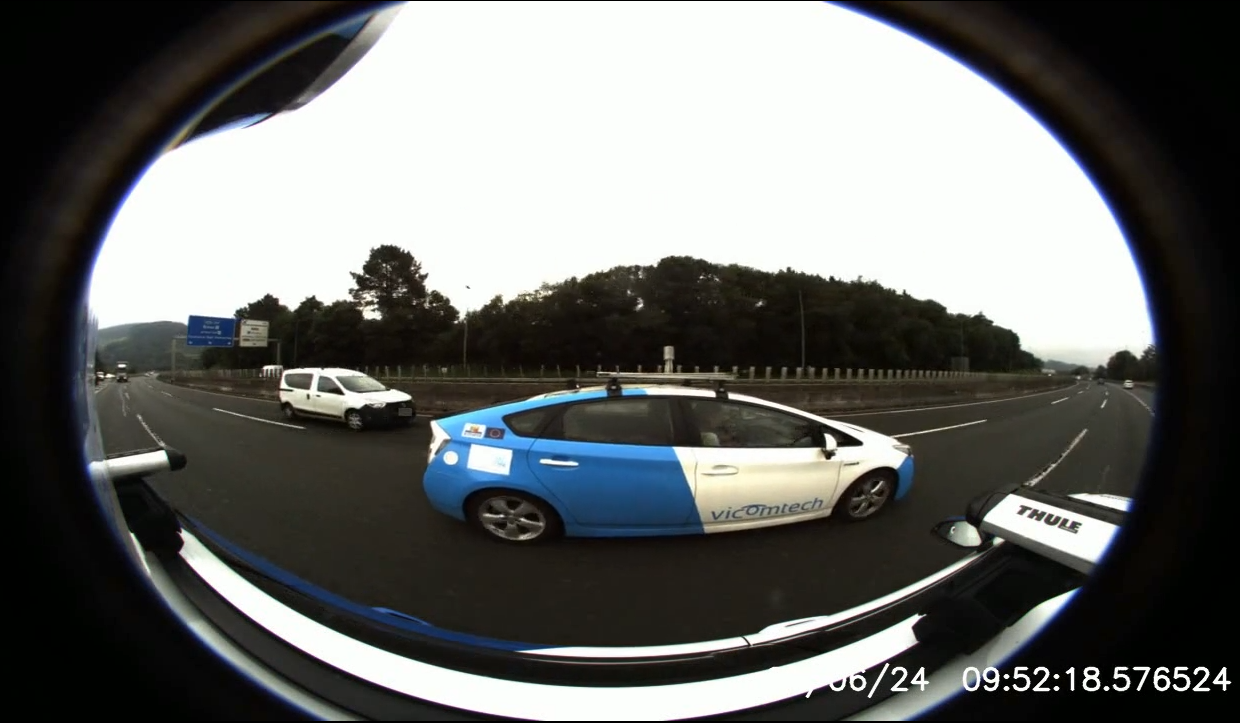}}
    \caption{Images captured during open road testing.}
\label{fig:tests_subfig}
\end{figure}

The tests were conducted using both the vehicle equipped with the CP system and the vehicle with AVL's DGT system. Both vehicles were driven along the BCC for over six hours, including both peak and off-peak traffic periods, with the road remaining open to regular traffic. During the tests, we kept the vehicles in close proximity, driving parallel to each other in adjacent lanes when traffic conditions permitted (see Figure \ref{fig:tests_subfig}). This approach aimed to ensure that both vehicles experienced similar scenes, facilitating a fair comparison between the two setups.

All systems in the architecture, including the vehicle with the CP system, the ground truth generator, and the BCC infrastructure, were synchronized using GNSS time synchronization, which has demonstrated a nanosecond-level of synchronization accuracy~\cite{dana1990role}, making it appropriate for precise latency evaluations in this application.
 
Our tests opted to transmit the CPMs at the maximum allowed frequency (10 Hz)~\cite{etsi2019intelligent}. This decision was based on the current low volume of V2X message traffic in the BCC and the desire to collect more data for a more precise evaluation of the perception and communication systems.

\subsection{DATA POST-PROCESSING}
\label{sec:eval:processing}
The ground truth data from the DGT system was generated in ASAM Open Simulation Interface (OSI) format and then parsed into CPMs. Both the ground truth and vehicle detections were generated at 10 Hz (every 100 ms), and a time offset of 50 ms was used to align the ground truth and detection frames. The position of the ground truth objects was then linearly interpolated to match the exact timestamp of the detection frame, using the velocity information provided in the ground truth data. Since the positions of the objects in the ground truth and detections were defined in separate coordinate systems (one for each vehicle), the GNSS coordinates of both vehicles in each frame were used to transform the ground truth object poses into the coordinate system of the connected vehicle. While efforts were made to keep both vehicles as close as possible during testing, frames where the distance between the two vehicles exceeded 50 meters were discarded for the perception evaluation, in order to ensure that both perception systems capture comparable scenarios and minimize discrepancies in the recorded data. 

During the experiments, the RSUs in the BCC corridor recorded all CPMs transmitted by nearby vehicles. However, the BCC infrastructure at that time lacked the capability to handle and process CPMs in real time. Additionally, we did not have access to integrate our LDM into the BCC infrastructure. To address these limitations, we devised an alternative approach. Network packets captured at the RSUs were employed to measure the latency between the OBUs and the RSUs. Then, we deployed our LDM on internal servers and replayed the captured network data to emulate real-time CPM reception within the infrastructure. This setup enabled the computation of the latency associated with processing CPMs and inserting the resulting data into the LDM, effectively simulating its functionality within the BCC corridor infrastructure. While the measured latency on our internal servers may differ from that of an actual edge deployment (e.g., due to potentially lower compute power at the edge), these results still provide a useful reference for an approximate system performance.

\subsection{EVALUATION METRICS}
\subsubsection{Communication Latencies}
In a CP system, the end-to-end latency of a CPM (from sensor capture at the sender to the CPM data fusion at the receiver) is the accumulation of multiple processing and transmission delays. Following the CPM delay chain model proposed in previous works (\cite{pilz2021components, pilz2023collective}), these latencies are categorized into three main components: sensing latency, communication latency, and fusion latency. Hence, the total CPM delay is expressed as in Equation~\eqref{eq:total_latency}:

\begin{equation}
    t_{\text{CPM}} = t_{\text{sensing}} + t_{\text{communication}} + t_{\text{fusion}}.
    \label{eq:total_latency}
\end{equation}

The sensing delay refers to the time required for sensors to capture data, process it, and detect relevant objects. It consists of three key subcomponents: the sensor data delay, cycle time and object detection delay. Therefore, it can be expressed as in Equation~\eqref{eq:sensing}:

\begin{equation}
    t_{\text{sensing}} = t_{\text{sensor\_data}} + t_{\text{cycle\_time}} + t_{\text{object\_detection}}.
    \label{eq:sensing}
\end{equation}

The communication delay consists of several components. On the transmission side, delay is introduced by buffering (CPMs cannot be sent immediately, as standards impose a maximum frequency of 10 Hz to prevent channel congestion) and packing (where CPMs are formatted into transmittable bitstreams). Similarly, on the receiver side, delays also occur due to CPM unpacking and distribution (the latter also considers CPM buffering before dissemination). Moreover, message transmission itself also introduces a delay between the transmitter and the receiver. The communication delay can thus be calculated using Equation~\eqref{eq:communication}:

\begin{equation}
    \begin{aligned}
        t_{\text{communication}} =\ & t_{\text{buffering}} + t_{\text{packing}} + t_{\text{transmission}} \\
        & + t_{\text{unpacking}} + t_{\text{distribution}}.
    \end{aligned}
    \label{eq:communication}
\end{equation}

Finally, the fusion delay includes matching, tracking, and high-level object data fusion at both the sender (local fusion) and receiver (global fusion). While local fusion is generally intended to combine multiple sensors, in our approach, it involves performing multi-object tracking on the LIDAR detections and synchronizing them with GNSS data to generate the CPMs. At the receiver, global fusion aggregates CPMs received from multiple vehicles into a single LDM, providing a coherent situational representation across multiple sources. 
Therefore, the fusion delay can be expressed as in Equation~\eqref{eq:fusion}:

\begin{equation}
    \begin{aligned}
        t_{\text{fusion}} =\ & t_{\text{local\_sender}} + t_{\text{global\_receiver}} \\
    \end{aligned}
    \label{eq:fusion}
\end{equation}

Table \ref{tab:delay_chain} illustrates the chronological order of each delay component in the end-to-end delay chain, from the sensor data delay in the transmitter to the global fusion at the receiver. 

\begin{table*}[ht]
    \caption{Diagram of the end-to-end CPM delay chain. Each delay component is ordered chronologically, from left to right.}
    \centering
    \begin{tabular}{|l|c| c |c|c |c|c| c |c|c |c|}
        \hline
         \textbf{Location} &\multicolumn{6}{c|}{Transmitter (vehicle)} & V2X channel & \multicolumn{3}{c|}{Receiver (RSU)} \\
        \hline
        \textbf{Component} &\multicolumn{3}{c|}{Sensing} & Fusion & \multicolumn{5}{c|}{Communication} & Fusion\\
        \hline
        \textbf{Subcomponent}& $t_{\text{sensor\_data}}$ & $t_{\text{cycle\_time}}$ & $t_{\text{object\_detection}}$ & $t_{\text{local\_sender}}$  & $t_{\text{buffering}}$ & $t_{\text{packing}}$ & $t_{\text{transmission}}$ &  $t_{\text{unpacking}}$ & $t_{\text{distribution}}$ & $t_{\text{global\_receiver}}$ \\
        \hline
    \end{tabular}
    \label{tab:delay_chain}
\end{table*}

The overall latency of the ICP architecture was evaluated using several metrics: average and median latencies, standard deviation, 95th percentile and 99th percentile. Both end-to-end latencies and the latencies of individual modules within the ICP architecture were measured. Each of these values was compared to those reported in the state of the art.

\subsubsection{Communication Ranges}
To evaluate the communication range between vehicles and the BCC infrastructure, the Packet Delivery Ratio (PDR) was used as a performance metric. For the V2I direction, the PDR is measured by comparing network captures at the vehicle's transmission point with the packets received at each RSU, and calculating the proportion of successfully received CPMs. For I2V, the PDR is measured similarly (i.e., as the fraction of successfully received messages at the vehicle relative to the RSUs' transmissions) but using CAMs, as the RSUs did not transmit CPMs.  Although CAMs are usually smaller than CPMs, they provide a reasonable approximation of I2V PDR trends, as packet loss is primarily driven by distance-dependent propagation rather than message size. By analysing the PDR for different distance ranges, its degradation can be observed as the distance between the vehicle and each RSU increases. 

\subsubsection{Perception System}
The perception system was evaluated using precision, recall, accuracy, and F1-score, which are widely used metrics in object detection \cite{padilla2021comparative}. To associate predictions with ground truth data and determine true detections, a threshold was applied to the 3D Generalized Intersection over Union (GIoU) value between predicted and ground truth 3D cuboids \cite{xu20193d}. Following previous works \cite{pang2022simpletrack}, the threshold was set to -0.5.

\begin{figure}
    \centering
\includegraphics[width=1.0\linewidth]{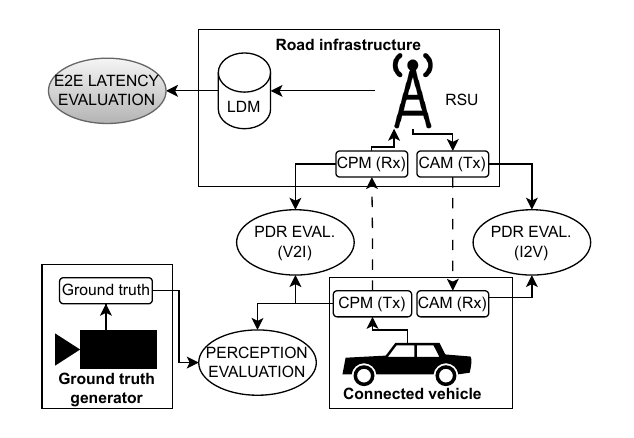}
    \caption{Block diagram summarizing the entire data flow used for evaluation. The shaded block (end-to-end latency evaluation) indicates that part of it is directly measured, while other subcomponents (sensing data, cycle time) are derived from manufacturer specifications. All other blocks are fully measured.}
    \label{fig:eval_diagram}
\end{figure}

Figure \ref{fig:eval_diagram} provides an overview of the complete data flow in this work, from data generation to the evaluation of the previously defined metrics.

\section{RESULTS}
\label{sec:results}
It should be noted that most measurements reported in this paper correspond to CPMs in the V2I direction, rather than I2V or V2V. Although many components are shared across these directions, CP systems on the infrastructure side often rely on stationary sensors and different equipment setups, while vehicle antennas are usually mounted lower than RSUs. These factors may introduce variations compared to the values reported in this paper. Therefore, the presented results can serve as a general reference for CP capabilities and limitations, while recognising that some differences may exist in I2V and V2V applications.


\subsection{COMMUNICATION LATENCIES}
\label{sec:results:lat}

\begin{table}
    \caption{Average end-to-end and component-wise delays of the CP system, as defined in \cite{pilz2021components}, measured in this work. 
    In the communication component, two values are reported in our measurements corresponding to the results obtained with synchronous and asynchronous CPM transmission approaches, respectively. The values are compared with the estimated and measured delays reported in \cite{pilz2023collective}. All delays are given in milliseconds.}
    \centering
    \begin{tabular}{|l| >{\centering\arraybackslash}p{2.0cm} |>{\centering\arraybackslash}p{2.0cm} | c |}
        \hline
        \textbf{Component} &  \textbf{Estimated \cite{pilz2023collective} } & \textbf{Measured \cite{pilz2023collective} } & \textbf{Ours} \\
        \hline
        Sensing  & 206-280 & 207.7&  282.1 \\
        Communication  & 11-254 & 19.1-419.1 & 8.1 / 55.5 \\
        Fusion  & 4-6 & 104.8 &  9.1 \\
        \hline
        Total & 221-540 & 331.6-731.6 & 299.3 / 346.7   \\
        \hline
    \end{tabular}
    \label{tab:latencies_all}
\end{table}

Table \ref{tab:latencies_all} presents the measured average end-to-end latencies, along with the latencies of each individual component, and compares them with the estimated and measured delays reported by Pilz et al. \cite{pilz2023collective}. For the communication component, two values are reported, corresponding to cases where CPM transmission was either synchronized or not with the local perception system (as further explained in Section \ref{sec:results-delays-comm}).

The results indicate that the end-to-end delays observed in our system, as well as the delays obtained in each component, are consistent with those found in the state of the art.

The following subsections provide a more detailed analysis of each delay component, breaking them down into their respective subcomponents.

\subsubsection{Sensing Delay}

On the one hand, both sensor data delay and cycle time depend on the specific sensor used for perception. In this study, we use a Velodyne HDL-32E LiDAR operating at 10Hz, although it supports a maximum frequency of 20Hz. The upper bound of the sensor data delay can be estimated as the inverse of the sensor's maximum frequency, resulting in a delay of up to 50 ms. Regarding the cycle time, given the 10Hz operating frequency, the upper bound of the cycle time delay is approximately 100ms. These are the only subcomponents or delays not determined by empirical measurement in this study but rather derived from manufacturer specifications.

On the other hand, the object detection delay represents the time required to analyze sensor data and identify relevant objects before they are included in a CPM for transmission. During testing, the 3D detection model showed an average latency of 132.1 ms in the connected vehicle.

Table~\ref{tab:sensing_delay} presents the delay values of each sensing delay subcomponent in this study. Among them, object detection shows the highest average latency and is also the only delay independent of the sensing hardware. This makes it the component with the greatest potential for delay optimization.

\begin{table}
    \caption{Latencies of individual sensing delay subcomponents, in milliseconds.  Further details on sensing data and cycle time calculation (*) are provided in Section \ref{sec:results:lat}}.
    \centering
    \begin{tabular}{|l| c | c | c | c | c|}
        \hline
        \textbf{Component} & \textbf{Mean} & \textbf{Median} & \textbf{Std. Dev.} & \textbf{95\%} & \textbf{99\%}  \\
        \hline
        Sensing data* & 50.0 & - & - &- &-  \\
        Cycle time* & 100.0 & - & -& -  &-  \\
        Object detection & 132.1 & 130.0 & 10.0 & 144.5 & 157.7 \\
        \hline
    \end{tabular}
    \label{tab:sensing_delay}
\end{table}

\subsubsection{Communication Delay}
\label{sec:results-delays-comm}

\begin{table*}
    \caption{Latencies of individual communication delay subcomponents, in milliseconds. In buffering + packing, two values are reported corresponding to the results obtained with synchronous and asynchronous CPM transmission approaches, respectively.}
    \centering
    \begin{tabular}{|l| c | c | c | c | c | c|}
        \hline
        \textbf{Component} & \textbf{Mean} & \textbf{Median} & \textbf{Std. Dev.}  & \textbf{95\%} & \textbf{99\%}  \\
        \hline
        Buffering + packing & 5.6 / 53.0 & 5.5 / 49.5 & 1.2 / 23.6  & 8.7 / 99.7 & 11.9 / 108.9\\
        Transmission & 1.3 & 1.3 & 0.2 & 1.5 & 1.8 \\
        Unpacking  & 0.8 & 0.8 & 0.4 & 1.3 & 1.6\\
        Distribution  & 0.4  & 0.3  & 0.5  & 0.7  & 2.9 \\
        \hline
    \end{tabular}
    \label{tab:Communication_latency}
\end{table*}

Table \ref{tab:Communication_latency} provides a detailed breakdown of the delay introduced by each component contributing to the overall communication latency. The buffering and packing of the CPM are grouped together because they are handled within the same process. This is due to the use of a commercial OBU (Cohda Wireless MK5), where both operations are managed internally by the device’s software. The dual buffering and packing delay was measured as the time difference between the perception system output (that is, after object detection and local fusion) and the transmission of the CPM by the OBU.

This combined buffering and packaging process within the OBU introduces the highest latency among all components and accounts for the majority of the total communication delay. This process showed an average delay of 53 ms during the tests, with a high standard deviation and maximum values exceeding 100 ms. This is primarily due to the asynchronous transmission of CPMs at a fixed rate of 10 Hz, independent of the perception system frequency. In the worst case, this results in a buffering delay of up to 100 ms between perception and transmission. 

To analyse the impact of synchronizing the local perception system with message transmission, additional tests were conducted where CPMs were transmitted as soon as data became available to the OBU (synchronously), rather than at a fixed rate. This approach reduced the average delay to 5.6 ms, with much more stable latency (e.g., a 99th percentile of 11.9 ms), highlighting the importance of synchronization in minimizing unnecessary buffering delays. However, even when using a synchronous approach, compliance with CPM generation rules must be ensured to maintain the maximum transmission rate of 10 Hz (requiring a minimum interval of 100 ms between consecutive CPMs), regardless of the perception system's output \cite{etsi2019intelligent}.

The transmission delay between the OBU and RSUs remains consistently low, averaging around 1 ms per CPM, with minimal variation across different RSUs, as shown in Table~\ref{tab:OBU_RSU_transmission_latencies}. This stability is partly due to the low channel load during the tests and aligns well with the expected performance of ITS-G5 technology, confirming its reliability for low-latency V2X communication.

Both the unpacking and distribution processes demonstrated consistently low latency, with average delays below 1 ms. The unpacking step, which converts ASN.1 streams into a usable data structure, remained well under 1 ms on average. Similarly, during distribution, the unpacked CPM information is fed directly into the global fusion pipeline,  ensuring minimal delay. In both cases, latency remains stable, as reflected by consistently low values even at the 95th and 99th percentiles.

\begin{table}
    \caption{OBU-to-RSU transmission delay metrics, in milliseconds.}
    \centering
    \begin{tabular}{|c| c | c | c | c | c  |}
        \hline
        \textbf{RSU ID} & \textbf{Mean} & \textbf{Median} & \textbf{Std. Dev.}  & \textbf{95\%} & \textbf{99\%}  \\
        \hline
        501A & 1.28 & 1.26 & 0.14 & 1.51 & 1.76   \\
        502A & 1.29 & 1.26 & 0.21 & 1.54 & 1.85  \\
        503D & 1.30 & 1.26 & 0.25 & 1.54 & 1.91   \\
        504D & 1.28 & 1.25 & 0.15 & 1.52 & 1.72  \\
        505A & 1.26 & 1.24 & 0.11 & 1.48 & 1.62  \\
        506D & 1.30 & 1.28 & 0.14 & 1.53 & 1.69  \\
        507D & 1.31 & 1.27 & 0.29 & 1.53 & 2.19  \\
        508A & 1.31 & 1.28 & 0.15 & 1.53 & 1.87  \\
        510D & 1.32 & 1.29 & 0.21 & 1.58 & 1.87  \\
        101A & 1.29 & 1.27 & 0.11 & 1.52 & 1.64  \\
        102A & 1.32 & 1.29 & 0.16 & 1.56 & 1.87  \\
        103A & 1.29 & 1.27 & 0.12 & 1.50 & 1.68  \\
        104A & 1.39 & 1.31 & 1.35 & 1.58 & 1.98   \\
        106A & 1.31 & 1.29 & 0.16 & 1.54 & 1.70  \\
        107D & 1.32 & 1.30 & 0.12 & 1.54 & 1.67  \\
        108D & 1.32 & 1.30 & 0.13 & 1.57 & 1.70  \\
        109A & 1.32 & 1.29 & 0.20 & 1.46 & 2.34  \\
        111D & 1.32 & 1.29 & 0.14 & 1.57 & 1.85  \\
        112A & 1.31 & 1.28 & 0.19 & 1.55 & 1.93 \\
        114D & 1.31 & 1.28 & 0.17 & 1.54 & 1.89 \\
        \hline
        Total & 1.31 & 1.28 & 0.25  & 1.53 & 1.84  \\
        \hline
    \end{tabular}
    \label{tab:OBU_RSU_transmission_latencies}
\end{table}

\subsubsection{Fusion Latency}

\begin{table}
    \caption{Latencies of individual fusion delay subcomponents, in milliseconds.}
    \centering
    \begin{tabular}{|l| c | c | c | c | c | c|}
        \hline
        \textbf{Component} & \textbf{Mean} & \textbf{Median} & \textbf{Std. Dev.}  & \textbf{95\%} & \textbf{99\%}  \\
        \hline
        Sender fusion & 0.4  & 0.4  & 0.2   & 0.7  & 0.9 \\
        Receiver fusion  & 8.7 & 7.0 & 8.4 & 11.2  & 56.7 \\
        \hline
    \end{tabular}
    \label{tab:fusion_latency}
\end{table}

Table \ref{tab:fusion_latency} shows the breakdown of the delay introduced by each component that contributes to the overall fusion delay. The local fusion performed at the transmitting vehicle exhibits minimal latency, as it has been optimized in C++ for efficiency, making it significantly faster than previously analyzed components. The total perception latency on the transmitter side is given by the sum of the sensing delay and the local fusion delay (282.5 ms). On the receiver side, the global fusion (which also includes writing the fused information into the edge LDM) presents higher latency than local fusion but remains lower than that of earlier components.


\subsection{COMMUNICATION RANGES}
\label{sec:results-ranges}

\begin{figure}
    \centering
    \includegraphics[width=\linewidth]{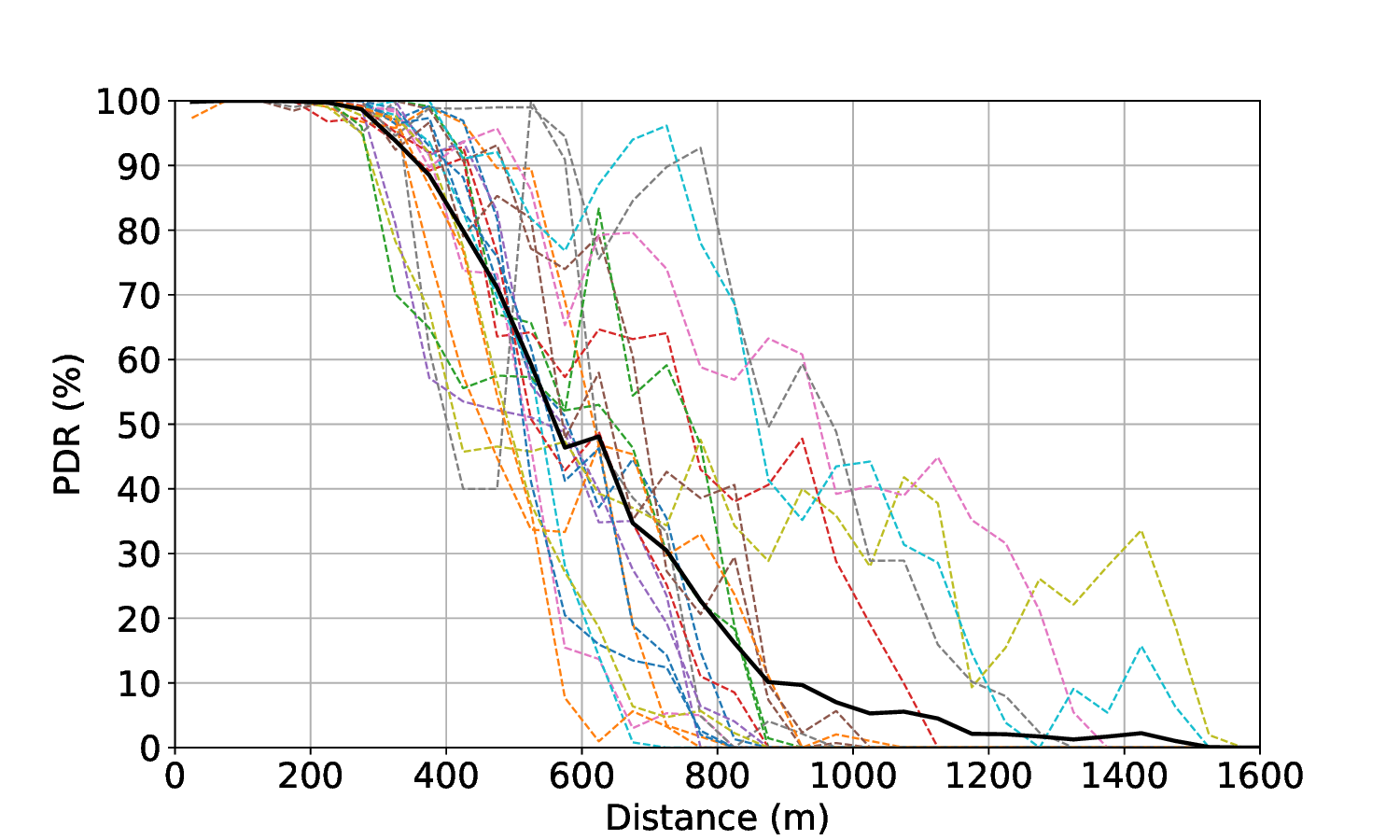}
    \caption{PDR (V2I) as a function of distance. Each dashed line represents an individual RSU, while the solid black line indicates the overall PDR across all RSUs.}
    \label{fig:com_range}
\end{figure}

Figure \ref{fig:com_range} illustrates the PDR (in the V2I direction) of the transmitted CPMs as a function of the distance between the vehicle at the time of transmission and each RSU. Each dashed colored line represents the PDR for an individual RSU, while the solid black line indicates the average across all RSUs. The PDR values were computed in 50-meter distance bins. As expected, greater communication distances resulted in higher packet loss, reducing the reliability of received CPMs. However, the PDR remained above 90\% up to 350 meters, as shown in Table~\ref{tab:range}, despite the challenging BCC terrain described in Section \ref{sec:architecture:bcc}. Measurements in the I2V direction (RSUs to vehicle) showed a similar trend with slightly lower values. This difference is likely due to hardware characteristics, as the vehicle's Cohda Wireless OBU has a higher transmit power than the Kapsch RIS-9260 RSUs (22 dBm vs 20 dBm, according to their datasheets), which leads to better PDR results in the V2I direction. Overall, the obtained results are consistent with the ones reported in the literature \cite{Maglogiannis2022}.

\begin{table}
    \caption{Average PDR across all RSUs for 50-meter distance bins. The V2I column reports PDR from the vehicle to RSUs, while the I2V column reports PDR from RSUs to the vehicle.}
    \centering
    \begin{tabular}{|c| c | c|}
        \hline
        \textbf{Distance (m)} & \textbf{PDR (\%), V2I} & \textbf{PDR (\%), I2V} \\
        \hline
        0 - 50 & 99.83 & 98.90 \\
        50 - 100 & 100.00 &98.25 \\
        100 - 150  & 100.00&98.24 \\
        150 - 200  & 99.91 &98.30\\
        200 - 250  & 99.71 &97.05 \\
        250 - 300  & 98.70&94.17 \\
        300 - 350  & 93.83 &88.37\\
        350 - 400  & 88.51 &76.05\\
        400 - 450  & 79.91 &66.85\\
        450 - 500  & 71.13 &51.09\\
        500 - 550  & 59.28 &47.70\\
        550 - 600  & 46.38 & 35.63\\
        600 - 650  & 48.09 & 28.75\\
        650 - 700  & 34.72 & 20.90\\
        700 - 750  & 30.46 & 16.95\\
        750 - 800  & 22.71 &11.76\\
        800 - 850  & 16.17 &10.12\\
        850 - 900  & 10.13 &6.82\\
        900 - 950  & 9.67  & 8.37\\
        950 - 1000  & 7.01 &5.29\\
        \hline
    \end{tabular}
    \label{tab:range}
\end{table}

\subsection{PERCEPTION SYSTEM}

\begin{figure}
    \centering
    \includegraphics[width=\linewidth]{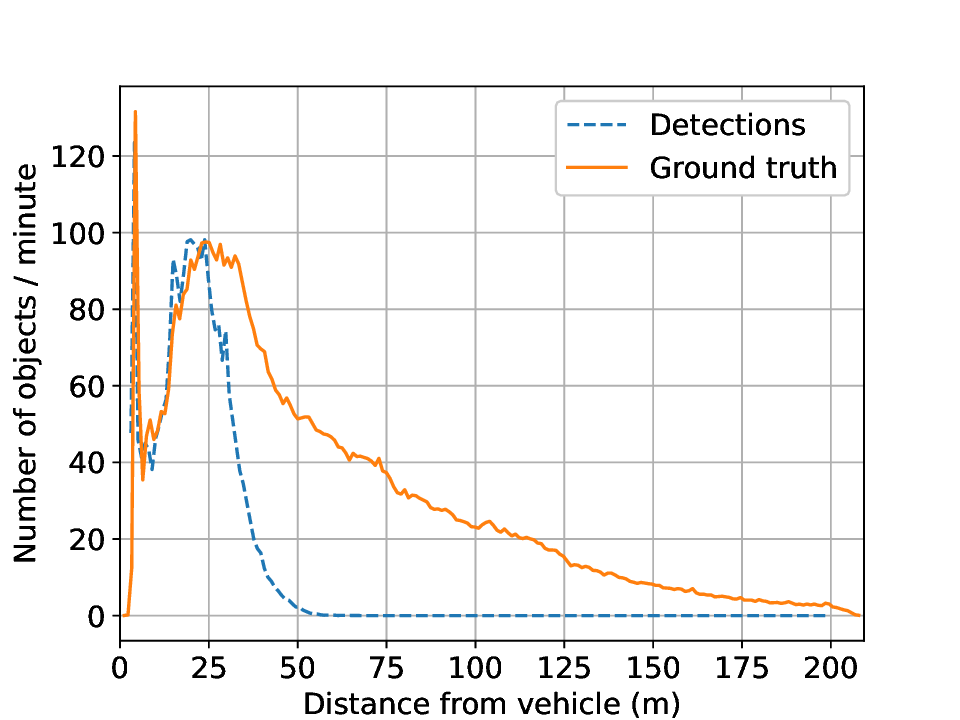}
    \caption{Distribution of the number of detected and ground truth objects per minute as a function of distance. The data is grouped into 1-meter intervals.}
    \label{fig:dets-distribution}
\end{figure}

\begin{figure}
    \centering
    \includegraphics[width=\linewidth]{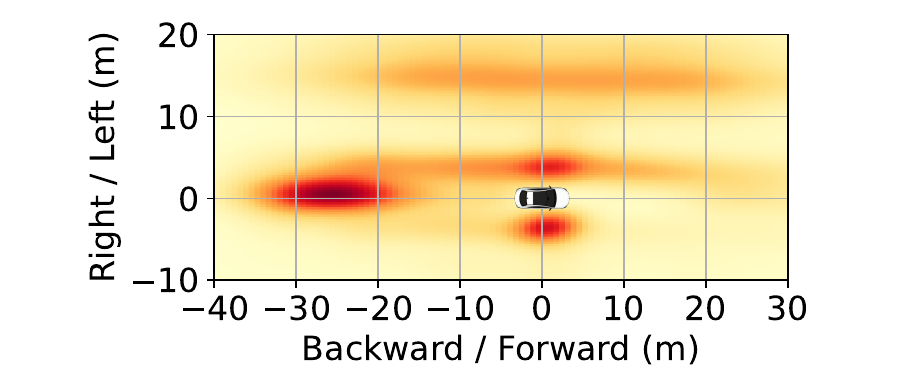}
    \caption{Heatmap of the accumulation of detected objects around the ego vehicle, in the local vehicle frame, during the whole test campaign. The darker the color, the higher the density.}
    \label{fig:heatmap}
\end{figure}

Figure \ref{fig:dets-distribution} shows how the objects detected by the connected vehicle and ground truth are distributed as a function of distance to each object. In both cases, the graphs show two prominent peaks: the first, around 5 meters,  corresponds to instances where the vehicles traveled in parallel in adjacent lanes; the second, around 25 meters, reflects the longitudinal distance maintained between the vehicles when the road configuration prevented parallel driving, as it can be seen more clearly in the heatmap in Figure \ref{fig:heatmap}. The density peak in the adjacent left lane (positive vertical axis values in the heatmap) is more extended than on the right (negative vertical axis), and can be attributed to vehicles overtaking the ego vehicle, as it was driven slightly slower than surrounding traffic (median speed of 80.7 km/h vs 83.5 km/h for other vehicles). Additionally, the heatmap shows a higher density 15 meters to the left of the ego vehicle (top of the heatmap), which likely corresponds to detections from oncoming traffic lanes. Beyond these distances, the number of detections by the connected vehicle decreases exponentially, with virtually no detections beyond 60 meters. The ground truth detections also decline with distance but at a slower rate. This reduction in ground truth detections may not only result from sensor range limitations but also from the curvy nature of the road, which obstructs the line of sight to objects at greater distances.

During the road tests, the main detected object classes were cars, motorbikes and trucks. Other road users, such as buses, appeared much less frequently and were excluded from this study due to their limited statistical significance.

\begin{table}
    \caption{3D object detection metrics for the detected classes and across different distance ranges.}
    \centering
    \begin{tabular}{|l| 
    >{\centering\arraybackslash}p{0.9cm}|
    c |
    >{\centering\arraybackslash}p{0.9cm}|
    c |
    >{\centering\arraybackslash}p{0.9cm}|}
        \hline
        \textbf{Class} & \textbf{Range (m)} & \textbf{Precision} & \textbf{Recall} & \textbf{Accuracy} & \textbf{F1-score} \\
        \hline
        \multirow{4}{*}{Motorbike} 
         & 0-30 & 0.239  & 0.082 & 0.065 & 0.122\\
          \cline{2-6}
         & 30-50 & 0.148 & 0.006 & 0.006  & 0.012 \\
         \cline{2-6}
         & 50-100 & 0.000 & 0.000 & 0.000 & 0.000 \\
         \cline{2-6}
         & Overall & 0.236 & 0.055 & 0.047 & 0.089\\
         \hline
        \multirow{4}{*}{Car} 
         & 0-30 & 0.579 & 0.657 & 0.445 & 0.615\\
          \cline{2-6}
         & 30-50 & 0.487 & 0.205 &  0.169 &  0.288\\
         \cline{2-6}
         & 50-100 & 0.274 & 0.001 & 0.001 & 0.002 \\
         \cline{2-6}
         & Overall & 0.562 & 0.334 & 0.265 & 0.419\\
        \hline
        \multirow{4}{*}{Truck} 
         & 0-30 & 0.386 & 0.249 & 0.178 & 0.303\\
          \cline{2-6}
         & 30-50 & 0.477 & 0.135 &  0.118 & 0.210 \\
         \cline{2-6}
         & 50-100 & 0.506 & 0.005 & 0.005 & 0.011 \\
         \cline{2-6}
         & Overall & 0.413 & 0.105 & 0.092 & 0.168\\
        \hline
    \end{tabular}
    \label{tab:perception_metrics}
\end{table}

We follow the approach from previous works (\cite{yu2022dair,xu2023v2v4real}) to analyse perception metrics across different distance ranges: 0-30 meters, 30-50 meters and 50-100 meters. The obtained perception metrics are presented in Table \ref{tab:perception_metrics}. As distance increases, recall shows a more pronounced degradation than precision, indicating that false negatives become more frequent with distance, while the rate of false positives remains comparatively more stable. While the cited prior works perform a similar analysis, direct metric comparisons are challenging for several reasons. First, they used a simplified 2D IoU in BEV to associate predictions with ground truth, whereas we use 3D GIoU, which provides a more informative measure for spatial overlap and localization in 3D space. Additionally, their studies focus on urban and suburban datasets, while ours is based on highway and high-speed scenarios. Furthermore, all our ground truth annotations are based on a highly calibrated sensor, distinct from the detection setup, while prior works use raw data from the detection setup for ground truth annotation as well, and therefore do not fully account for GNSS positioning and calibration errors of the detection setup itself.

\begin{figure}
    \centering
    \subfigure[Precision.]{\includegraphics[width=0.9\linewidth]{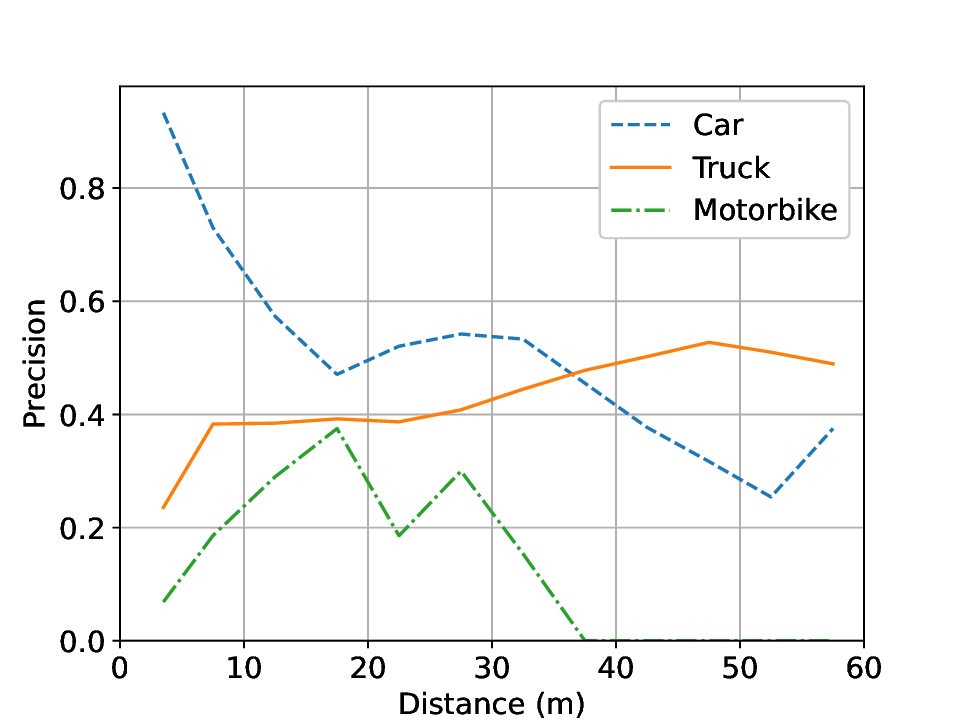}}\hfil
    \subfigure[Recall.]
{\includegraphics[width=0.9\linewidth]{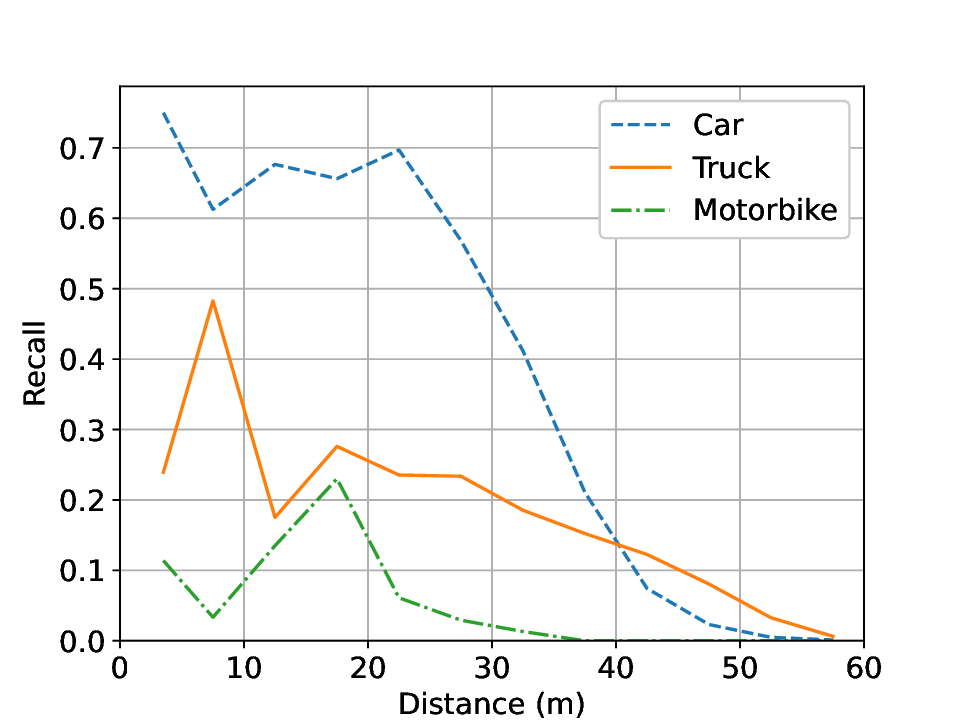}}
    \caption{3D object detection metrics as a function of distance to the ego vehicle.}
    \label{fig:perception-distribution}
\end{figure}

A deeper insight into the 3D object detection can be gained from Figure \ref{fig:perception-distribution}, which illustrates the precision and recall metrics of each object class as a function of the distance to the ego vehicle. The 0-60 meters range is divided into 5-meter bins, and the mentioned metrics are calculated inside each bin. The results show that recall is significantly more affected by distance than precision. Moreover, detection in larger objects (e.g., trucks) is more robust over increasing distances, whereas smaller objects are more sensitive to distance-related performance degradation.

\section{CONCLUSIONS}
\label{sec:conclusions}
The methodology proposed in this paper to evaluate a collective perception system provided valuable insights into the challenges and constraints of CPS in high-speed scenarios like highways. The observed decline in the F1-score with increasing distance is primarily driven by an increase in false negatives rather than a rise in false positives. Detection beyond 60 meters is almost non-existent with the setup employed in the connected vehicle in this study. At a speed of 106 Km/h, this corresponds to 2 seconds of travel time, posing a significant limitation for perception-based applications in high-speed environments. This finding highlights the increasing difficulty of detecting distant objects, particularly smaller ones. On the other hand, the relatively stable precision KPI suggests that, while distant objects are not always detected, the detections that do occur are generally reliable.

While the detection range could be partially extended with higher-resolution LiDAR sensors or multi-sensor fusion data techniques, such approaches come with increased deployment costs, higher computational demands (e.g., additional GPUs for multi-sensor inference), and additional processing delays due to late sensor fusion. Nevertheless, the observed decrease in detections with distance is not only a limitation of the connected vehicle’s setup but also a characteristic of the DGT system's data as well - despite its use of a multi-sensor system, advanced data fusion, and processing on high-performance servers without the physical constraints of onboard vehicle computers. This suggests that the challenge extends beyond specific sensor configurations, and may be an inherent limitation of onboard perception systems and CPS services, influenced by factors such as sensor range, road curvature, and occlusions limiting line-of-sight. 

This limitation highlights the importance of collective perception in high-speed environments like highways, where vehicles and infrastructure can exchange perception data to overcome individual perception constraints. The ITS-G5 communication ranges measured in Section \ref{sec:results} are significantly higher than the analysed individual perception ranges, indicating that V2X can greatly enhance the overall situational awareness of the vehicle. While roadside sensors enabling Collective Perception Services should be strategically placed near critical areas that require enhanced coverage (such as highway on-ramps or congestion-prone zones), the extended communication range of ITS-G5 allows RSUs to be positioned more flexibly, reducing deployment constraints. As future work, an algorithm can be developed to optimize the placement of new RSU deployments, ensuring maximum coverage and efficiency while considering existing infrastructure. 
Infrastructure-assisted collective perception in these critical areas could provide earlier and more reliable awareness of surrounding traffic, leading to enhanced state estimation and reduced uncertainty in occluded areas, as well as supporting more informed motion planning and control decisions. These improvements would not only increase safety but also driving comfort, as CAVs would have additional time to adjust their trajectories, minimizing abrupt maneuvers. Future studies should also extend the evaluation to more common, lower-cost vehicle setups, such as camera-based detectors or non-inertial localization systems, and analyse the performance degradation compared to advanced systems like the one used in this paper.

Regarding the communication latencies measured in this study, they generally align with the values reported in the literature. However, these latencies become particularly critical in high-speed scenarios, as even short delays translate into significant distances traveled by vehicles. This spatial discrepancy can significantly impact the accuracy and reliability of cooperative perception. 
Therefore, data fusion techniques in such environments could benefit from delayed multi-agent state estimation, using predictive models to estimate the current positions based on their last known states and measured latencies. By incorporating such temporal compensation strategies, CP systems can ensure robust situational awareness despite communication and processing delays.

Two key components contribute significantly to the overall end-to-end latency in the collective perception: the perception module and the buffering delay introduced by the OBU during the CPM transmission. 

On the one hand, the perception system represents a major source of latency, as Deep Learning-based object detection models require substantial computational resources. A potential optimization strategy involves exploring efficient model inference techniques, such as TensorRT and ONNX-based acceleration, which could reduce processing times. Previous studies have evaluated the latency improvements that can be achieved through such optimizations \cite{verma2021performance} and assessed their impact on the system performance \cite{shafique2023deep}. These optimizations could be explored as part of future work in the current study to further enhance the efficiency of the collective perception system.

On the other hand, an asynchronous CPM transmission system can introduce a buffering delay of up to 100 ms in the worst case, increasing the baseline end-to-end delay in a synchronized approach (299 ms) by approximately 33\%. To mitigate this issue, it is crucial to synchronize the CPM transmission with the output of the perception system. This synchronization must be carefully managed to comply with the ETSI standards that establish that two consecutive CPMs cannot be transmitted in less than 100 ms. In a synchronous approach, there may still be instances where a newly detected object cannot be immediately transmitted if a CPM has already been sent within the last 100 ms, thus introducing additional latency to the collective perception. However, this drawback is not exclusive to synchronous transmission, as the same issue can arise in asynchronous approaches. By implementing adaptive CPM generation rules rather than relying on a fixed transmission rate, the number of CPMs sent per second can be reduced, minimizing instances where critical detections are delayed due to the 100 ms window constraint. Therefore, an optimal solution should combine synchronous CPM transmission with optimized CPM generation rules, reducing avoidable latency and improving the responsiveness of CP systems while ensuring compliance with regulatory standards.

\section*{Acknowledgments}
This project has been funded by the Provincial Council of Bizkaia (Bizkaiko Foru Aldundia) as part of the Promotion of Innovation in Road Infrastructure 2024. We would like to express our gratitude to AVL Deutschland GmbH for their support and assistance in this work.

%

\bibliographystyle{IEEEtran}
\bibliography{references}

@article{ma2021perception,
  title={{Perception entropy: A metric for multiple sensors configuration evaluation and design}},
  author={Ma, Tao and Liu, Zhizheng and Li, Yikang},
  journal={arXiv preprint arXiv:2104.06615},
  year={2021}
}

@inproceedings{hu2022investigating,
  title={{Investigating the impact of multi-lidar placement on object detection for autonomous driving}},
  author={Hu, Hanjiang and Liu, Zuxin and Chitlangia, Sharad and Agnihotri, Akhil and Zhao, Ding},
  booktitle={Proceedings of the IEEE/CVF conference on computer vision and pattern recognition},
  pages={2550--2559},
  year={2022}
}

@inproceedings{pilz2021components,
  title={{The components of cooperative perception-a proposal for future works}},
  author={Pilz, Christoph and Ulbel, Andrea and Steinbauer-Wagner, Gerald},
  booktitle={2021 IEEE International Intelligent Transportation Systems Conference (ITSC)},
  pages={7--14},
  year={2021},
  organization={IEEE}
}

@article{pilz2023collective,
  title={Collective perception: A delay evaluation with a short discussion on channel load},
  author={Pilz, Christoph and Sammer, Peter and Piri, Esa and Grossschedl, Udo and Steinbauer-Wagner, Gerald and Kuschnig, Lukas and Steinberger, Alina and Schratter, Markus},
  journal={IEEE Open Journal of Intelligent Transportation Systems},
  year={2023},
  publisher={IEEE}
}

@inproceedings{schiegg2023accounting,
  title={Accounting for the special role of infrastructure-assisted collective perception},
  author={Schiegg, Florian Alexander and Rueeck, Anna--Lisa and Gamerdinger, J{\"o}rg and Tchouankem, Hugues and Xhoxhi, Edmir and Volk, Georg},
  booktitle={2023 IEEE 26th International Conference on Intelligent Transportation Systems (ITSC)},
  pages={189--195},
  year={2023},
  organization={IEEE}
}

@inproceedings{wolff2023infrastructure,
  title={{Infrastructure-Assisted Collective Perception Service with Emphasis on Vulnerable Road User Perception}},
  author={Wolff, Vincent Albert},
  booktitle={International Conference on Smart Cities and Green ICT Systems},
  pages={197--211},
  year={2023},
  organization={Springer}
}

@inproceedings{chtourou2021collective,
  title={Collective perception service for connected vehicles and roadside infrastructure},
  author={Chtourou, Ameni and Merdrignac, Pierre and Shagdar, Oyunchimeg},
  booktitle={2021 IEEE 93rd Vehicular Technology Conference (VTC2021-Spring)},
  pages={1--5},
  year={2021},
  organization={IEEE}
}

@inproceedings{volk2021towards,
  title={Towards realistic evaluation of collective perception for connected and automated driving},
  author={Volk, Georg and Delooz, Quentin and Schiegg, Florian A and Von Bernuth, Alexander and Festag, Andreas and Bringmann, Oliver},
  booktitle={2021 IEEE International Intelligent Transportation Systems Conference (ITSC)},
  pages={1049--1056},
  year={2021},
  organization={IEEE}
}

@inproceedings{huang2019performance,
  title={Performance modelling of v2v based collective perceptions in connected and autonomous vehicles},
  author={Huang, Hui and Fang, Wenqi and Li, Huiyun},
  booktitle={2019 IEEE 44th Conference on Local Computer Networks (LCN)},
  pages={356--363},
  year={2019},
  organization={IEEE}
}

@article{dana1990role,
  title={The role of GPS in precise time and frequency dissemination},
  author={Dana, Peter H and Penrod, Bruce M},
  journal={GPS World},
  volume={1},
  number={4},
  pages={38--43},
  year={1990},
  publisher={Beuth Verlag Berlin}
}

@inproceedings{pang2022simpletrack,
  title={Simpletrack: Understanding and rethinking 3d multi-object tracking},
  author={Pang, Ziqi and Li, Zhichao and Wang, Naiyan},
  booktitle={European Conference on Computer Vision},
  pages={680--696},
  year={2022},
  organization={Springer}
}

@inproceedings{yu2022dair,
  title={Dair-v2x: A large-scale dataset for vehicle-infrastructure cooperative 3d object detection},
  author={Yu, Haibao and Luo, Yizhen and Shu, Mao and Huo, Yiyi and Yang, Zebang and Shi, Yifeng and Guo, Zhenglong and Li, Hanyu and Hu, Xing and Yuan, Jirui and others},
  booktitle={Proceedings of the IEEE/CVF Conference on Computer Vision and Pattern Recognition},
  pages={21361--21370},
  year={2022}
}

@inproceedings{xu2023v2v4real,
  title={V2v4real: A real-world large-scale dataset for vehicle-to-vehicle cooperative perception},
  author={Xu, Runsheng and Xia, Xin and Li, Jinlong and Li, Hanzhao and Zhang, Shuo and Tu, Zhengzhong and Meng, Zonglin and Xiang, Hao and Dong, Xiaoyu and Song, Rui and others},
  booktitle={Proceedings of the IEEE/CVF Conference on Computer Vision and Pattern Recognition},
  pages={13712--13722},
  year={2023}
}

@article{etsi2019intelligent,
  title={Intelligent transport system (its); vehicular communications; basic set of applications; analysis of the collective-perception service (cps)},
  author={ETSI, ITS},
  journal={Draft TR 103 562 V0. 0.15},
  year={2019}
}

@inproceedings{verma2021performance,
  title={Performance evaluation of deep learning compilers for edge inference},
  author={Verma, Gaurav and Gupta, Yashi and Malik, Abid M and Chapman, Barbara},
  booktitle={2021 IEEE international parallel and distributed processing symposium workshops (IPDPSW)},
  pages={858--865},
  year={2021},
  organization={IEEE}
}

@article{shafique2023deep,
  title={{Deep Learning Performance Characterization on GPUs for Various Quantization Frameworks}},
  author={Shafique, Muhammad Ali and Munir, Arslan and Kong, Joonho},
  journal={AI},
  volume={4},
  number={4},
  pages={926--948},
  year={2023},
  publisher={MDPI}
}

@article{car2019guidance,
  title={Guidance for day 2 and beyond roadmap},
  author={Car 2 Car Communication Consortium and others},
  journal={Car2Car Communication Consortium, C2CCC WP},
  volume={2072},
  year={2019}
}

@inproceedings{almeida2023real,
  title={On the real evaluation of a collective perception service},
  author={Almeida, Pedro and Figueiredo, Andreia and Rito, Pedro and Lu{\'\i}s, Miguel and Sargento, Susana},
  booktitle={NOMS 2023-2023 IEEE/IFIP Network Operations and Management Symposium},
  pages={1--6},
  year={2023},
  organization={IEEE}
}

@inproceedings{merwaday2021infrastructure,
  title={Infrastructure assisted efficient collective perception service for connected vehicles},
  author={Merwaday, Arvind and Jha, Satish C and Sivanesan, Kathiravetpillai and Alvarez, Ignacio J and Baltar, Leonardo Gomes and Banjade, Vesh Raj Sharma and Sehra, Suman A},
  booktitle={2021 IEEE Vehicular Networking Conference (VNC)},
  pages={119--120},
  year={2021},
  organization={IEEE}
}

@inproceedings{zimmer2024tumtraf,
  title={Tumtraf v2x cooperative perception dataset},
  author={Zimmer, Walter and Wardana, Gerhard Arya and Sritharan, Suren and Zhou, Xingcheng and Song, Rui and Knoll, Alois C},
  booktitle={Proceedings of the IEEE/CVF conference on computer vision and pattern recognition},
  pages={22668--22677},
  year={2024}
}

@article{yuan2022keypoints,
  title={Keypoints-based deep feature fusion for cooperative vehicle detection of autonomous driving},
  author={Yuan, Yunshuang and Cheng, Hao and Sester, Monika},
  journal={IEEE Robotics and Automation Letters},
  volume={7},
  number={2},
  pages={3054--3061},
  year={2022},
  publisher={IEEE}
}

@inproceedings{chen2019f,
  title={F-cooper: Feature based cooperative perception for autonomous vehicle edge computing system using 3D point clouds},
  author={Chen, Qi and Ma, Xu and Tang, Sihai and Guo, Jingda and Yang, Qing and Fu, Song},
  booktitle={Proceedings of the 4th ACM/IEEE Symposium on Edge Computing},
  pages={88--100},
  year={2019}
}

@book{iso2013road,
   title={Road vehicles--vehicle dynamics and road-holding ability--vocabulary},
  author={ISO, D and others},
  year={2013},
  publisher={Beuth Berlin}
}

@ARTICLE{zfernandez2023,
  author={Fernández, Zaloa and Martín, Angel and Pérez, Josu and García, Mikel and Velez, Gorka and Murciano, Federico and Peters, Sebastian},
  journal={IEEE Access}, 
  title={{Challenges and Solutions for Service Continuity in Inter-PLMN Handover for Vehicular Applications}}, 
  year={2023},
  volume={11},
  number={},
  pages={8904-8919},
  doi={10.1109/ACCESS.2023.3239694}}

@INPROCEEDINGS{via2022,
  author={Vía, Selva and Payaró, Miquel and Eckert, Kurt and Mühleisen, Maciej and Wendt, Stefan and Fischer, Edwin and Hetzer, Dirk and Kousaridas, Apostolos},
  booktitle={2022 IEEE Future Networks World Forum (FNWF)}, 
  title={{5GCroCo: Key 5G technologies and trial results for seamless cross-border CAM services (Invited Paper)}}, 
  year={2022},
  volume={},
  number={},
  pages={24-27},
  doi={10.1109/FNWF55208.2022.00013}}

@INPROCEEDINGS{slamnik2024,
  author={Slamnik-Kriještorac, Nina and Vandenberghe, Wim and Limani, Xhulio and Oostendorp, Eric and de Groote, Eva and Maglogiannis, Vasilis and Naudts, Dries and Schackmann, Peter-Paul and Kusumakar, Rakshith and Kural, Karel and Kia, Ghazaleh and Campodonico, Maria Chiara and Moerman, Ingrid and Marquez-Barja, Johann M.},
  booktitle={2024 Joint European Conference on Networks and Communications \& 6G Summit (EuCNC/6G Summit)}, 
  title={{5G-enhanced Teleoperation in Real-Life Port Environments: Lessons Learned from the 5G-Blueprint Project}}, 
  year={2024},
  volume={},
  number={},
  pages={973-978},
  doi={10.1109/EuCNC/6GSummit60053.2024.10597023}}

@INPROCEEDINGS{Gallego2024,
  author={Vázquez-Gallego, Francisco and Nasreddine, Jad and Codina, Marc and Cordero, Bruno and Carmona-Cejudo, Estela and Trullenque, Martín and Camps-Mur, Daniel and Murillo, Yuri and Seguret, Philippe and Polo, Javier and Luque, José López},
  booktitle={2024 IEEE 99th Vehicular Technology Conference (VTC2024-Spring)}, 
  title={{Performance Evaluation of 5G Standalone Seamless Home Routed Roaming for Connected Mobility in Cross-Border Scenarios}}, 
  year={2024},
  volume={},
  number={},
  pages={1-7},
  doi={10.1109/VTC2024-Spring62846.2024.10683417}}

@ARTICLE{Moradi2023,
  author={Moradi-Pari, Ehsan and Tian, Danyang and Bahramgiri, Mojtaba and Rajab, Samer and Bai, Sue},
  journal={IEEE Transactions on Intelligent Transportation Systems}, 
  title={{DSRC Versus LTE-V2X: Empirical Performance Analysis of Direct Vehicular Communication Technologies}}, 
  year={2023},
  volume={24},
  number={5},
  pages={4889-4903},
  doi={10.1109/TITS.2023.3247339}}

@article{Guerrieri2021,
author = {Guerrieri, Marco and Mauro, Raffaele and Pompigna, Andrea and Isaenko, Natalia},
title = {{Road Design Criteria and Capacity Estimation Based on Autonomous Vehicles Performances. First Results from the European C-Roads Platform and A22 Motorway}},
journal = {Transport and Telecommunication Journal},
number = {2},
volume = {22},
year = {2021},
pages = {230--243}
}

@Article{Kang2023,
AUTHOR = {Kang, Junhee and Tak, Sehyun and Park, Sungjin},
TITLE = {{Analyzing the Impact of C-ITS Services on Driving Behavior: A Case Study of the Daejeon–Sejong C-ITS Pilot Project in South Korea}},
JOURNAL = {{Sustainability}},
VOLUME = {15},
YEAR = {2023},
NUMBER = {16},
ARTICLE-NUMBER = {12655},
URL = {https://www.mdpi.com/2071-1050/15/16/12655},
ISSN = {2071-1050},
DOI = {10.3390/su151612655}
}

@article{thandavarayan2020generation,
  title={Generation of cooperative perception messages for connected and automated vehicles},
  author={Thandavarayan, Gokulnath and Sepulcre, Miguel and Gozalvez, Javier},
  journal={IEEE Transactions on Vehicular Technology},
  volume={69},
  number={12},
  pages={16336--16341},
  year={2020},
  publisher={IEEE}
}

@inproceedings{shule2022tracking,
  title={Tracking accuracy based generation rules of collective perception messages},
  author={Shule, LI and Wolff, Vincent Albert},
  booktitle={2022 IEEE 25th International Conference on Intelligent Transportation Systems (ITSC)},
  pages={4157--4162},
  year={2022},
  organization={IEEE}
}

@article{chtourou2021context,
  title={Context-aware content selection and message generation for collective perception services},
  author={Chtourou, Ameni and Merdrignac, Pierre and Shagdar, Oyunchimeg},
  journal={Electronics},
  volume={10},
  number={20},
  pages={2509},
  year={2021},
  publisher={MDPI}
}

@inproceedings{thandavarayan2019analysis,
  title={Analysis of message generation rules for collective perception in connected and automated driving},
  author={Thandavarayan, Gokulnath and Sepulcre, Miguel and Gozalvez, Javier},
  booktitle={2019 IEEE Intelligent Vehicles Symposium (IV)},
  pages={134--139},
  year={2019},
  organization={IEEE}
}

@INPROCEEDINGS{mgarcia2023,
  author={García, Mikel and Vélez, Gorka and Pérez, Josu and Martín, {\'A}ngel and Fernández, Zaloa and Aginako, Naiara},
  booktitle={2023 IEEE 26th International Conference on Intelligent Transportation Systems (ITSC)}, 
  title={{Edge Dynamic Map architecture for C-ITS applications}}, 
  year={2023},
  volume={},
  number={},
  pages={5060-5065},
  doi={10.1109/ITSC57777.2023.10422497}}

@ARTICLE{jarin2024,
  author={Arin, Javier and Velez, Gorka and Bustamante, Paul},
  journal={IEEE Access}, 
  title={{A C-ITS Architecture for MEC and Cloud Native Back-End Services}}, 
  year={2024},
  volume={12},
  number={},
  pages={64531-64550},
  doi={10.1109/ACCESS.2024.3397467}}

@ARTICLE{fmogollon2024,
  author={Mogollon, Felipe and Fernandez, Zaloa and Martin, Angel and Ortega, Juan Diego and Velez, Gorka},
  journal={IEEE Transactions on Intelligent Vehicles}, 
  title={Data streaming platform for crowd-sourced vehicle dataset generation}, 
  year={2024},
  volume={},
  number={},
  pages={1-10},
  doi={10.1109/TIV.2024.3486926}}

@ARTICLE{Zhengwei2024,
  author={Bai, Zhengwei and Wu, Guoyuan and Barth, Matthew J. and Liu, Yongkang and Akin Sisbot, Emrah and Oguchi, Kentaro and Huang, Zhitong},
  journal={IEEE Transactions on Intelligent Transportation Systems}, 
  title={{A Survey and Framework of Cooperative Perception: From Heterogeneous Singleton to Hierarchical Cooperation}}, 
  year={2024},
  volume={25},
  number={11},
  pages={15191-15209},
  doi={10.1109/TITS.2024.3436012}}

@article{etsi2011102,
  title={{102 863 (v1. 1.1): Intelligent transport systems (ITS); vehicular communications; basic set of applications; local dynamic map (LDM); rationale for and guidance on standardization}},
  author={ETSI, TR},
  journal={ETSI: Sophia Antipolis, France},
  year={2011}
}

@InProceedings{Yin_2021_CVPR,
    author    = {Yin, Tianwei and Zhou, Xingyi and Krahenbuhl, Philipp},
    title     = {{Center-Based 3D Object Detection and Tracking}},
    booktitle = {Proceedings of the IEEE/CVF Conference on Computer Vision and Pattern Recognition (CVPR)},
    month     = {June},
    year      = {2021},
    pages     = {11784-11793}
}

@InProceedings{Caesar_2020_CVPR,
author = {Caesar, Holger and Bankiti, Varun and Lang, Alex H. and Vora, Sourabh and Liong, Venice Erin and Xu, Qiang and Krishnan, Anush and Pan, Yu and Baldan, Giancarlo and Beijbom, Oscar},
title = {{nuScenes: A Multimodal Dataset for Autonomous Driving}},
booktitle = {Proceedings of the IEEE/CVF Conference on Computer Vision and Pattern Recognition (CVPR)},
month = {June},
year = {2020}
}

@ARTICLE{Maglogiannis2022,
  author={Maglogiannis, Vasilis and Naudts, Dries and Hadiwardoyo, Seilendria and van den Akker, Daniel and Marquez-Barja, Johann and Moerman, Ingrid},
  journal={IEEE Transactions on Network and Service Management}, 
  title={{Experimental V2X Evaluation for C-V2X and ITS-G5 Technologies in a Real-Life Highway Environment}}, 
  year={2022},
  volume={19},
  number={2},
  pages={1521-1538},
  doi={10.1109/TNSM.2021.3129348}}

@article{montero2023multi,
  title={Multi-camera BEV video-surveillance system for efficient monitoring of social distancing},
  author={Montero, David and Aranjuelo, Nerea and Leskovsky, Peter and Loyo, Est{\'\i}baliz and Nieto, Marcos and Aginako, Naiara},
  journal={Multimedia Tools and Applications},
  volume={82},
  number={22},
  pages={34995--35019},
  year={2023},
  publisher={Springer}
}

@article{xu20193d,
  title={3D-GIoU: 3D generalized intersection over union for object detection in point cloud},
  author={Xu, Jun and Ma, Yanxin and He, Songhua and Zhu, Jiahua},
  journal={Sensors},
  volume={19},
  number={19},
  pages={4093},
  year={2019},
  publisher={MDPI}
}

@article{padilla2021comparative,
  title={A comparative analysis of object detection metrics with a companion open-source toolkit},
  author={Padilla, Rafael and Passos, Wesley L and Dias, Thadeu LB and Netto, Sergio L and Da Silva, Eduardo AB},
  journal={Electronics},
  volume={10},
  number={3},
  pages={279},
  year={2021},
  publisher={MDPI}
}

@article{hajisami2022tutorial,
  title={A tutorial on the LTE-V2X direct communication},
  author={Hajisami, Abolfazl and Lansford, Jim and Dingankar, Aasif and Misener, Jim},
  journal={IEEE Open Journal of Vehicular Technology},
  volume={3},
  pages={388--398},
  year={2022},
  publisher={IEEE}
}

@article{masuda2022feature,
  title={Feature-based vehicle identification framework for optimization of collective perception messages in vehicular networks},
  author={Masuda, Hidetaka and El Marai, Oussama and Tsukada, Manabu and Taleb, Tarik and Esaki, Hiroshi},
  journal={{IEEE Transactions on Vehicular Technology}},
  volume={72},
  number={2},
  pages={2120--2129},
  year={2022},
  publisher={IEEE}
}

@article{ku2023uncertainty,
  title={Uncertainty-aware task offloading for multi-vehicle perception fusion over vehicular edge computing},
  author={Ku, Yu-Jen and Baidya, Sabur and Dey, Sujit},
  journal={IEEE Transactions on Vehicular Technology},
  volume={72},
  number={11},
  pages={14906--14923},
  year={2023},
  publisher={IEEE}
}

@ARTICLE{zhang2024efficient,
  author={Zhang, Jingyu and Yang, Kun and Wang, Hanqi and Sun, Peng and Song, Liang},
  journal={{IEEE Transactions on Vehicular Technology}}, 
  title={{Efficient Vehicular Collaborative Perception Based on Saptial-Temporal Feature Compression}}, 
  year={2024},
  volume={73},
  number={11},
  pages={16125-16133},
  doi={10.1109/TVT.2024.3403263}}

@article{jia2023mass,
  title={MASS: Mobility-aware sensor scheduling of cooperative perception for connected automated driving},
  author={Jia, Yukuan and Mao, Ruiqing and Sun, Yuxuan and Zhou, Sheng and Niu, Zhisheng},
  journal={IEEE Transactions on Vehicular Technology},
  volume={72},
  number={11},
  pages={14962--14977},
  year={2023},
  publisher={IEEE}
}

@article{malik2023collaborative,
  title={Collaborative perception—the missing piece in realizing fully autonomous driving},
  author={Malik, Sumbal and Khan, Muhammad Jalal and Khan, Manzoor Ahmed and El-Sayed, Hesham},
  journal={Sensors},
  volume={23},
  number={18},
  pages={7854},
  year={2023},
  publisher={MDPI}
}

@ARTICLE{Asabe2024,
  author={Asabe, Yu and Javanmardi, Ehsan and Nakazato, Jin and Tsukada, Manabu and Esaki, Hiroshi},
  journal={IEEE Open Journal of Vehicular Technology}, 
  title={Enhancing Reliability in Infrastructure-Based Collective Perception: A Dual-Channel Hybrid Delivery Approach With Real-Time Monitoring}, 
  year={2024},
  volume={5},
  number={},
  pages={1124-1138},
  doi={10.1109/OJVT.2024.3443877}}

@inproceedings{herrero2023pedestrian,
  title={Pedestrian movement prediction based on camera vision and deep learning model},
  author={Herrero, Alejandro Miranda and P{\'e}rez-Benito, Cristina and Salido, David Pujol and Gual, Joan Serrat and Jevti{\'c}, Aleksandar},
  booktitle={2023 IEEE 26th International Conference on Intelligent Transportation Systems (ITSC)},
  pages={2233--2238},
  year={2023},
  organization={IEEE}
}



 

\begin{IEEEbiography}[{\includegraphics[width=1in,height=1.25in,clip,keepaspectratio]{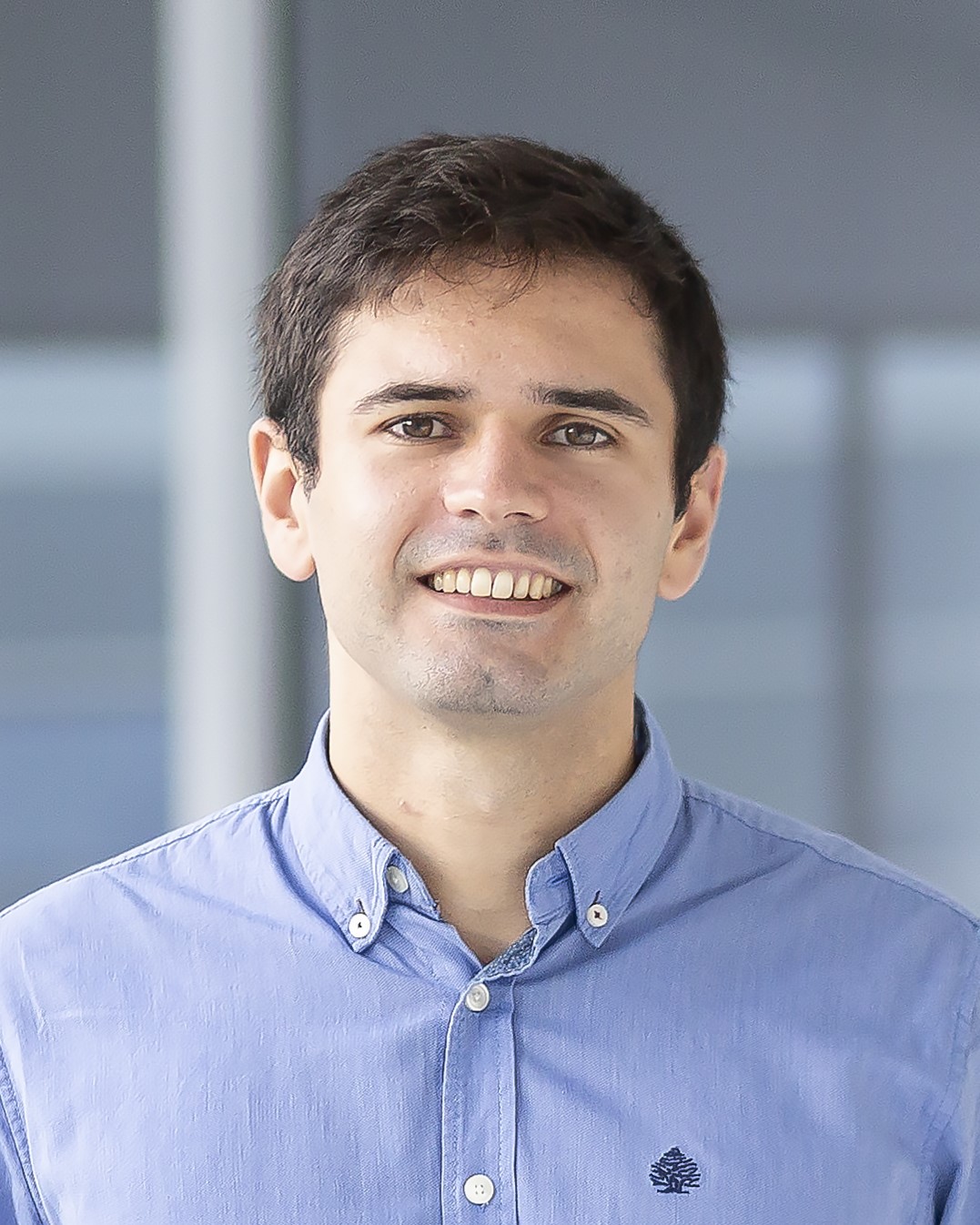}}]{Jon Ander Iñiguez de Gordoa} received his M.Sc. degree in Telecommunication Engineering from the University of Navarra, Spain, in 2021. He is currently a Ph.D. candidate at the Faculty of Informatics at the University of the Basque Country, Spain. He works as a researcher in the Connected, Cooperative and Automated Systems division in Vicomtech. His research interests include V2X communications, Artificial Intelligence, and simulation applied to the ITS sector.
\end{IEEEbiography}


\begin{IEEEbiography}[{\includegraphics[width=1in,height=1.25in,clip,keepaspectratio]{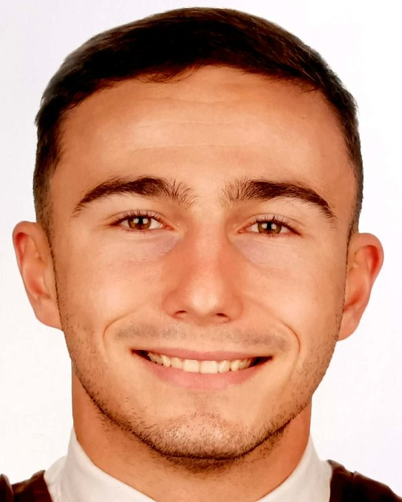}}]{Iker Alkorta} obtained his M.Sc. in Telecommunication Engineering from the University of the Basque Country, Spain, in 2024, and his Master's degree in Computer Engineering with a specialization in Internet of Things from the Illinois Institute of Technology, Chicago, in the same year. 
During his time at Vicomtech, he worked as a research assistant in the Connected, Cooperative, and Automated Systems department. His research interests are centered around cybersecurity, Internet of Things, V2X communications, and Artificial Intelligence.
\end{IEEEbiography}


\begin{IEEEbiography}[{\includegraphics[width=1in,height=1.25in,clip,keepaspectratio]{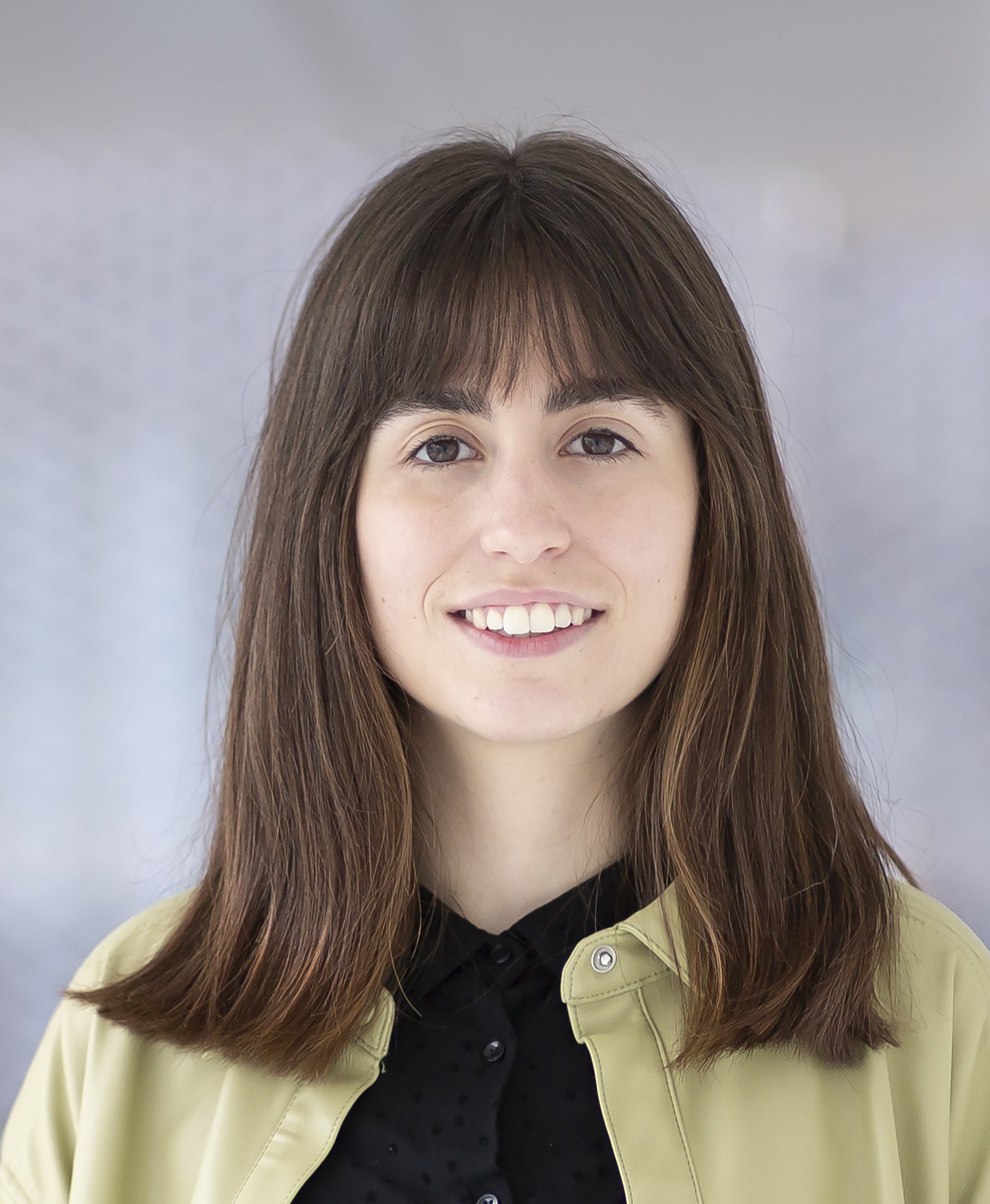}}]{Itziar Urbieta} earned a B.Sc. in Energy Engineering from Mondragon Unibertsitatea in 2017 and an M.Sc. in Transport Systems from the University of the Basque Country in 2018. Since 2018, she has been a researcher at Vicomtech in the Intelligent Systems for Mobility and Logistics department. She is currently a Ph.D. candidate at Deusto University in the Engineering for the Information Society and Sustainable Development program. Her research focuses on data-driven modeling, semantic tools, and simulations for the ITS sector.
\end{IEEEbiography}


\begin{IEEEbiography}[{\includegraphics[width=1in,height=1.25in,clip,keepaspectratio]{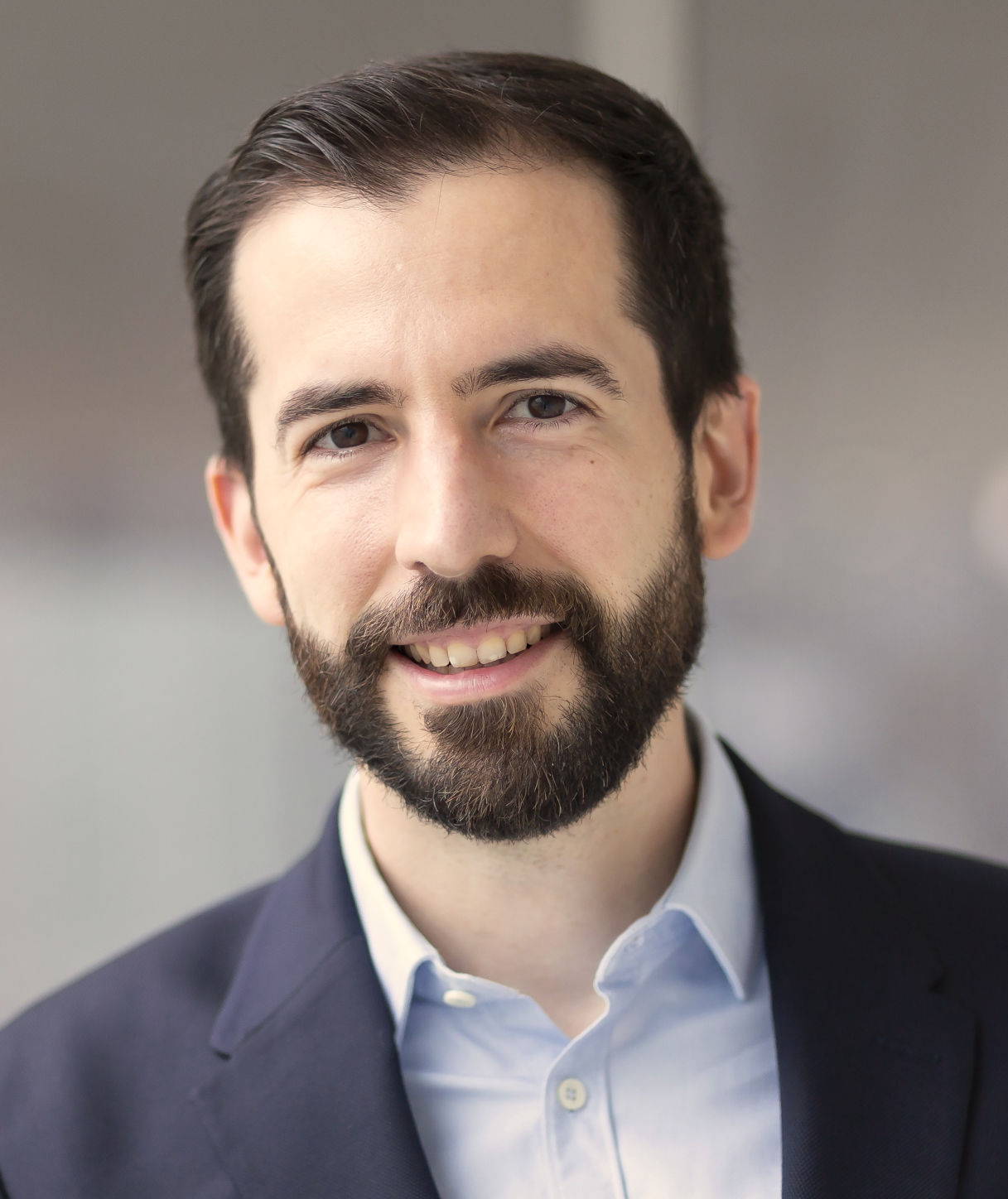}}]{Gorka Velez} received his M.Sc. degree in Electronic Engineering from the University of Mondragon, Spain, in 2007, and his Ph.D. degree from the University of Navarra, Spain, in 2012. He currently leads the Connected and Cooperative Situation Awareness research line within the Intelligent Transportation Systems (ITS) division at Vicomtech. His research interests include Internet of Things, Big Data, and Artificial Intelligence applied to the ITS sector.
\end{IEEEbiography}

\begin{IEEEbiography}[{\includegraphics[width=1in,height=1.25in,clip]{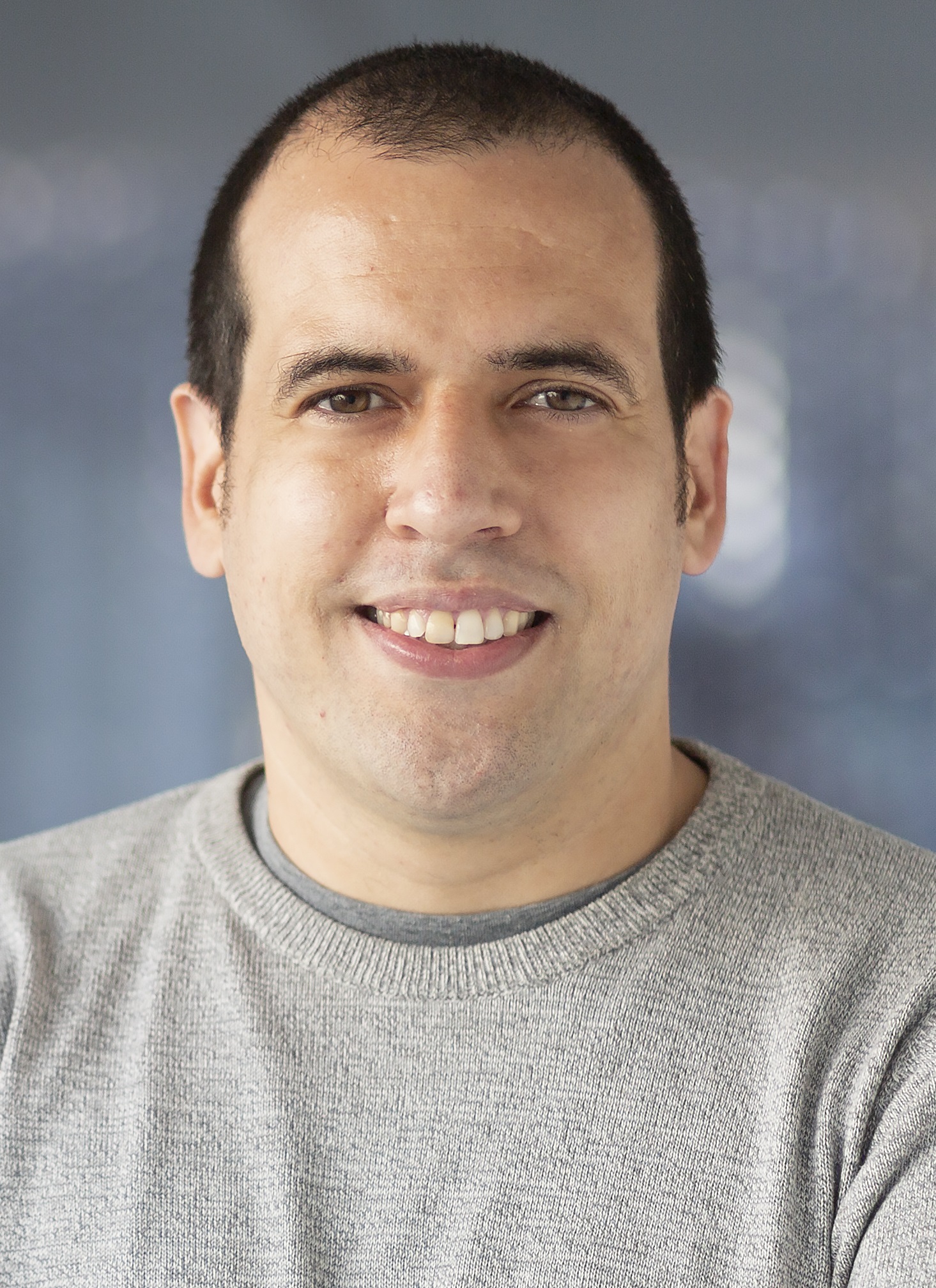}}]{Andoni Mujika}
holds a degree in Mathematics from the University of the Basque Country and a doctorate in Information and Communication Technologies from Rey Juan Carlos University. He is currently an assistant professor in the Department of Computer Science and Artificial Intelligence at the University of the Basque Country. Previously, he worked as a researcher in the Intelligent Transport Systems and Engineering department at the Vicomtech technology center. His research interests include the physical simulation of living organisms; visualization and interaction with point clouds; user-friendly human-computer interaction; and optimization algorithms for logistics and mobility.
\end{IEEEbiography}



\vfill

\end{document}